\documentclass[traditabstract]{aa} 
\usepackage{graphicx}
\usepackage{txfonts}
\usepackage{multirow}

\newcommand{\myrule}{\rule[-0.1cm]{0.cm}{0.5cm}} 
\newcommand\lsun{L_{\odot}}
\newcommand\msun{M_{\odot}}
\newcommand\rsun{R_{\odot}}
\newcommand\mjup{M_\mathrm{Jup}}

\newcommand\iso{ISO\,}
\newcommand\mperyr{M_{\odot}\,\mathrm{yr}^{-1}}

\begin{document}

\title{Discovery of an outflow of the very low-mass star \iso143\thanks{Based 
     on observations obtained at the Very Large Telescope of the 
	    European Southern Observatory at Paranal, Chile 
	    in program 
	    080.C-0904(A)    
	    082.C-0023(A+B)  
	  }}

 \titlerunning{Discovery of an outflow of the very low-mass star \iso143}

   \author{V. Joergens,\inst{1,2}
           T. Kopytova, \inst{2}
           A. Pohl\inst{2}
          }

   \institute{
        Universit\"at Heidelberg,
	Zentrum f\"ur Astronomie,
	Institut f\"ur Theoretische Astrophysik
	Albert-Ueberle-Str. 2
	69120 Heidelberg, Germany,
             \email{viki@mpia.de}
	\and     
	     Max-Planck Institut f\"ur Astronomie, 
             K\"onigstuhl~17, 69117 Heidelberg, Germany
             }

   \date{Received ; accepted Sept 27, 2012}

  \abstract
   {We discover that the very young very low-mass star \iso143 (M5)
is driving an outflow based on spectro-astrometry 
of forbidden [S\,II] emission lines at 6716\,{\AA} and 6731\,{\AA} observed in UVES/VLT spectra.
This adds another object to the handful of brown dwarfs and very low-mass stars 
(M5-M8) for which an outflow has been confirmed and which show that the T~Tauri phase continues
at very low masses.
We find the outflow of \iso143 to be intrinsically asymmetric 
and the accretion disk to not obscure the outflow,
as only the red outflow component is visible in the [S\,II] lines.
\iso143 is only the third T~Tauri object showing a
stronger red outflow component in spectro-astrometry, after RW\,Aur (G5) and \iso217 (M6.25).
We show here that, including \iso143,
two out of seven outflows confirmed in the very low-mass regime (M5-M8) are intrinsically 
asymmetric.
We measure a spatial extension of the outflow in [S\,II] of up to 200-300\,mas (about 30-50\,AU) and
velocities of up to 50-70\,km\,s$^{-1}$. 
We furthermore detect line emission of \iso143 in Ca\,II (8498\,{\AA}), 
O\,I (8446\,{\AA}), He\,I (7065\,{\AA}), 
and weakly in [Fe\,II] (7155\,{\AA}).
Based on a line profile analysis and decomposition we demonstrate that (i) 
the Ca\,II emission can be attributed to chromospheric activity, a variable wind, and the 
magnetospheric infall zone,
(ii) the O\,I emission mainly to accretion-related processes but also a wind,
and (iii) the He\,I emission to chromospheric or coronal activity.
We estimate a mass outflow rate of \iso143 of $\sim10^{-10}$\,$\mperyr$
and a mass accretion rate in the range of $\sim10^{-8}$ to
$\sim10^{-9}$\,$\mperyr$. These values are consistent 
with those of other brown dwarfs and very low-mass stars.
The derived $\dot{M}_{out}/\dot{M}_{acc}$ ratio of 1-20\,\% does 
not support previous findings of
this number being very large ($>$40\,\%) for very low-mass objects.
}


   \maketitle
%

\section{Introduction}
\label{sect:intro}

While many details of the origin of brown dwarfs are still unknown,
it was established in the last few years that during their early evolution brown dwarfs
resemble higher mass T~Tauri stars in main properties. 
Very young brown dwarfs (a few Myr) show surface activity, such as cool spots (e.g., Joergens et al. 2003).
There is evidence that brown dwarfs have disks from 
mid-infrared (mid-IR, e.g., Comer\'on et al. 2000; Jayawardhana et al. 2003; Luhman et al. 2008) and
far-IR/submm excess emission (Klein et al. 2003; Scholz et al. 2006; 
Harvey et al. 2012a, 2012b). Many of these disks have been found to 
be actively accreting (e.g., Mohanty et al. 2005; Herczeg \& Hillenbrand 2008; 
Bacciotti et al. 2011; Rigliaco et al. 2011), and
several show signs of grain growth and crystallization (e.g., Apai et al. 2005; Pascucci et al. 2009).
Furthermore, very young brown dwarfs rotate on average more slowly 
(e.g., Joergens \& Guenther 2001; Joergens et al. 2003; Caballero et al. 2004) 
than their older counterparts (e.g., Bailer-Jones \& Mundt 2001; Mohanty \& Basri 2003), 
which is indicative of a magnetic braking mechanism due to interaction with the disk.

Jets and outflows are known to be a by-product of accretion in the star-formation process,
and they have been observed for many classical T~Tauri stars 
(CTTS, e.g., Ray et al. 2007 for a review).
It has recently been shown that brown dwarfs and very low-mass stars (VLMS) 
are also able to drive T~Tauri-like outflows based on 
the detection of forbidden emission lines (FELs, e.g., Fernandez \& Comer\'on 2001;
Looper et al. 2010),
spectro-astrometry of FELs (Whelan et al. 2005, 2007, 2009a, 2009b;
Bacciotti et al. 2011; Joergens et al. 2012), 
and direct imaging in the CO J=2-1 transition (Phan-Bao et al. 2008).
To date, outflows have been confirmed for six
M5 to M8 type brown dwarfs and VLMS providing evidence that objects of a tenth of a 
solar mass to less than 30\,$\mjup$ 
can launch powerful outflows: Par-Lup3-4 (M5), $\rho$\,Oph\,102 
(M5.5\footnote{Spectral type $\rho$\,Oph\,102: K. Luhman, private communication.}), 
\iso217 (M6.25), LS-R\,CrA\,1 (M6.5), 2M1207 (M8), and ISO-Oph\,32 (M8).

We present here the discovery of an outflow of the VLMS \iso143 (M5) 
based on the spectro-astrometry of [S\,II] lines in high-resolution UVES/VLT spectra.
This adds another object to the small sample of detected outflows 
for brown dwarfs and VLMS.
We investigate the outflow properties of \iso143. Furthermore,
we show that \iso143 exhibits several other activity-related emission lines and 
localize their formation sites.

The paper is organized as follows:
After a summary of the known properties of \iso143 (Sect.\,\ref{sect:iso143}),
the acquisition and analysis of high-resolution UVES/VLT spectra
is described (Sect.\,\ref{sect:obs}). 
In Sect.\,\ref{sect:results}, 
the results of the emission line study and the spectro-astrometric detection
of the outflow is presented, followed by our conclusions on \iso143 in
Sect.\,\ref{sect:concl}.

\section{The very low-mass star \iso143}
\label{sect:iso143}

\iso143\footnote{Simbad name: \object{ISO-ChaI 143}}  
is a very low-mass star of spectral type M5 situated in the Cha\,I star-forming region
(Luhman 2007).
An estimate of its mass based on a comparison of
effective temperature and luminosity 
($T_\mathrm{eff}$=3125\,K, $L_\mathrm{bol}$=0.088\,$L_{\odot}$, Luhman 2007)
with evolutionary models (Baraffe et al. 1998) yields a value of 0.18\,$\msun$.
\iso143 exhibits mid- and far-IR excess emission providing
evidence of a circumstellar disk.
These detections were made with the InfraRed Array Camera (IRAC) on board the {\em Spitzer} satellite (3.6, 4.5, 5.8, 8.0\,$\mu$m),
the Multiband Imaging Photometer for Spitzer (MIPS, 24\,$\mu$m, Damjanov et al. 2007; Luhman et al. 2008),
and the Photoconductor Array Camera and Spectrometer (PACS) of the {\em Herschel} mission (70, 160\,$\mu$m, Harvey et al. 2012b).
An IR spectrum of \iso143 taken by the {\em Spitzer}/InfraRed Spectrograph (IRS, 5-36\,$\mu$m) shows 
a weak 10\,$\mu$m silicate feature (Manoj et al. 2011).
\iso 143 was classified as a class II source based on the spectral slope between 8\,$\mu$m  and 24\,$\mu$m 
(Luhman et al. 2008; Manoj et al. 2011). The disk appears to be relatively flat and to have undergone   
substantial dust settling (Manoj et al. 2011).
Modeling of its spectral energy distribution (SED),
including {\em Herschel}/PACS data at 70\,$\mu$m  and 160\,$\mu$m 
(Harvey et al. 2012b; Y. Liu, pers. comm.), yields 
a total disk mass of 5$\times 10^{-6} M_\odot$ (assuming a standard gas-to-dust ratio of 100),
a disk flaring index of less than 1.1,
and an inclination less than 60$^{\circ}$, with the most likely value being 
15$^{\circ}$-25$^{\circ}$, i.e. close to face on, although inclination values 0$^{\circ}$--50$^{\circ}$ 
are almost equally likely. 
\iso143 shows H$\alpha$ in emission with an equivalent width of 118\,{\AA} (Luhman 2004),
indicating that the disk is actively accreting.

\begin{table*}
\begin{minipage}[t]{\columnwidth}
\centering
\caption{
\label{tab:obslog} 
Observing log of \iso143.
}
\renewcommand{\footnoterule}{}  
\begin{tabular}{cccccc}
\hline
\hline
\myrule
Date & HJD & Exptime & Seeing   & Slit PA & RV             \\
     &     & [sec]   & [arcsec] & $[$deg] & [km\,s$^{-1}$] \\
\hline
\myrule
2008 03 19 & 2454544.69473 & 2$\times$1700 & 0.67 & 10.4\,$\pm$\,8.3 & 16.315\,$\pm$\,0.082 \\
2009 01 15 & 2454846.84705 & 2$\times$1700 & 0.55 & ~\,3.1\,$\pm$\,8.2 & 17.273\,$\pm$\,0.191 \\
2009 02 10 & 2454872.87430 & 2$\times$1700 & 0.62 & 42.2\,$\pm$\,7.8 & 16.986\,$\pm$\,0.398 \\
\hline
\end{tabular}
\tablefoot{
The listed entries are: observing date,
heliocentric julian date at the middle of the exposure, exposure time, the averaged seeing corrected by airmass,
slit position angle (PA), and radial velocity (RV, from Joergens, in prep.).
The slit PA was not kept fixed during observations, and each 2$\times$1700\,sec exposure samples a 
PA range of about $\pm$\,8\,deg, as listed.
}
\end{minipage}
\end{table*}


\section{High-resolution spectra and spectro-astrometry}
\label{sect:obs}

We observed \iso143 with the Ultraviolet and Visual Echelle Spectrograph (UVES, Dekker et al. 2000) 
attached to the VLT 8.2\,m KUEYEN telescope
in three nights in 2008 and 2009 in the red optical wavelength regime 
at a spectral resolution $\lambda$/$\Delta \lambda$ of 40\,000.  
The spatial sampling was 0.182$^{\prime\prime}$/pixel.
An observing log is given in Table\,\ref{tab:obslog}. 

\begin{figure*}
\vbox{
\hbox{\mbox{}
 \hfill\includegraphics[width=5.7cm,clip]{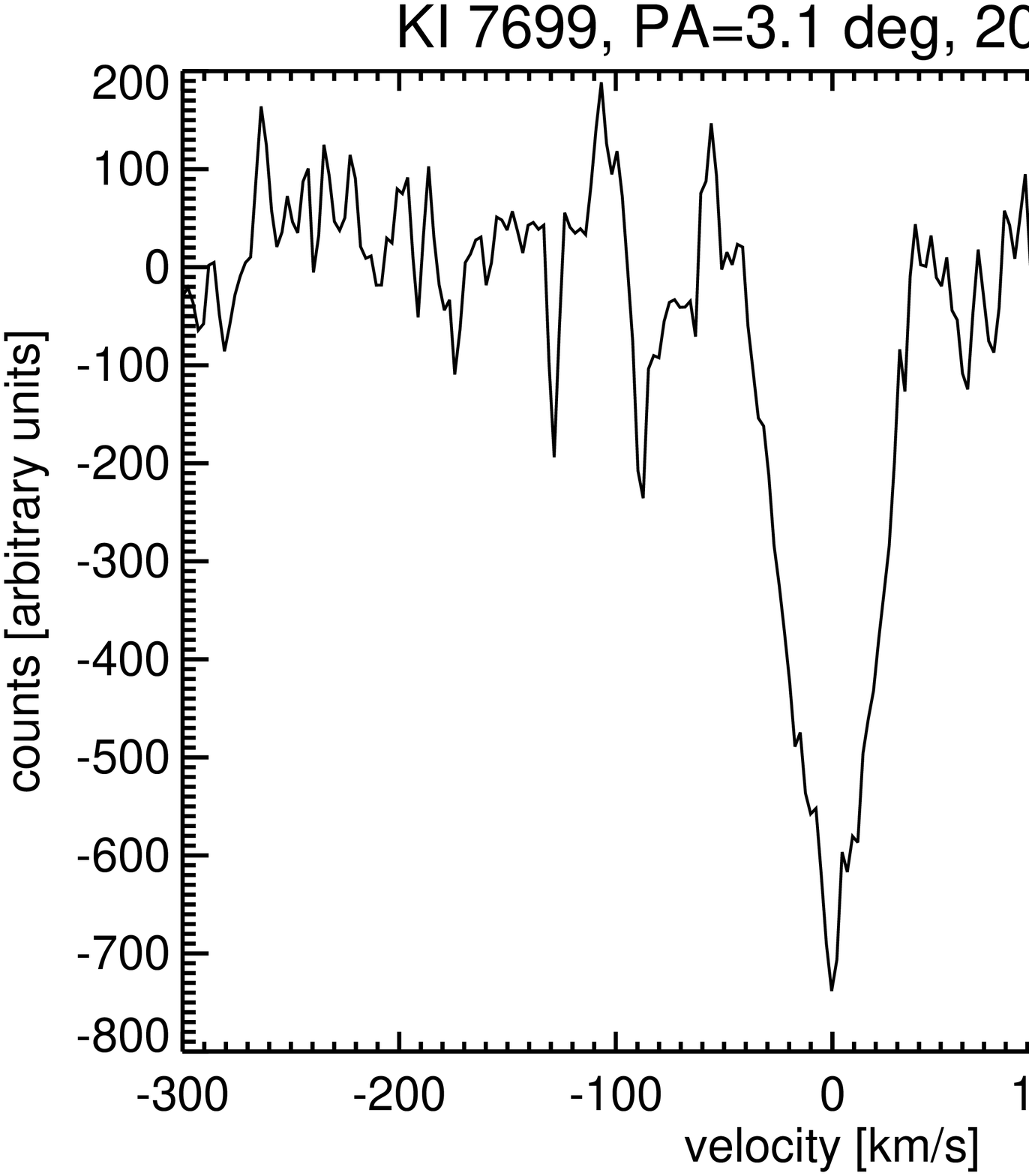}
 \hfill\includegraphics[width=5.7cm,clip]{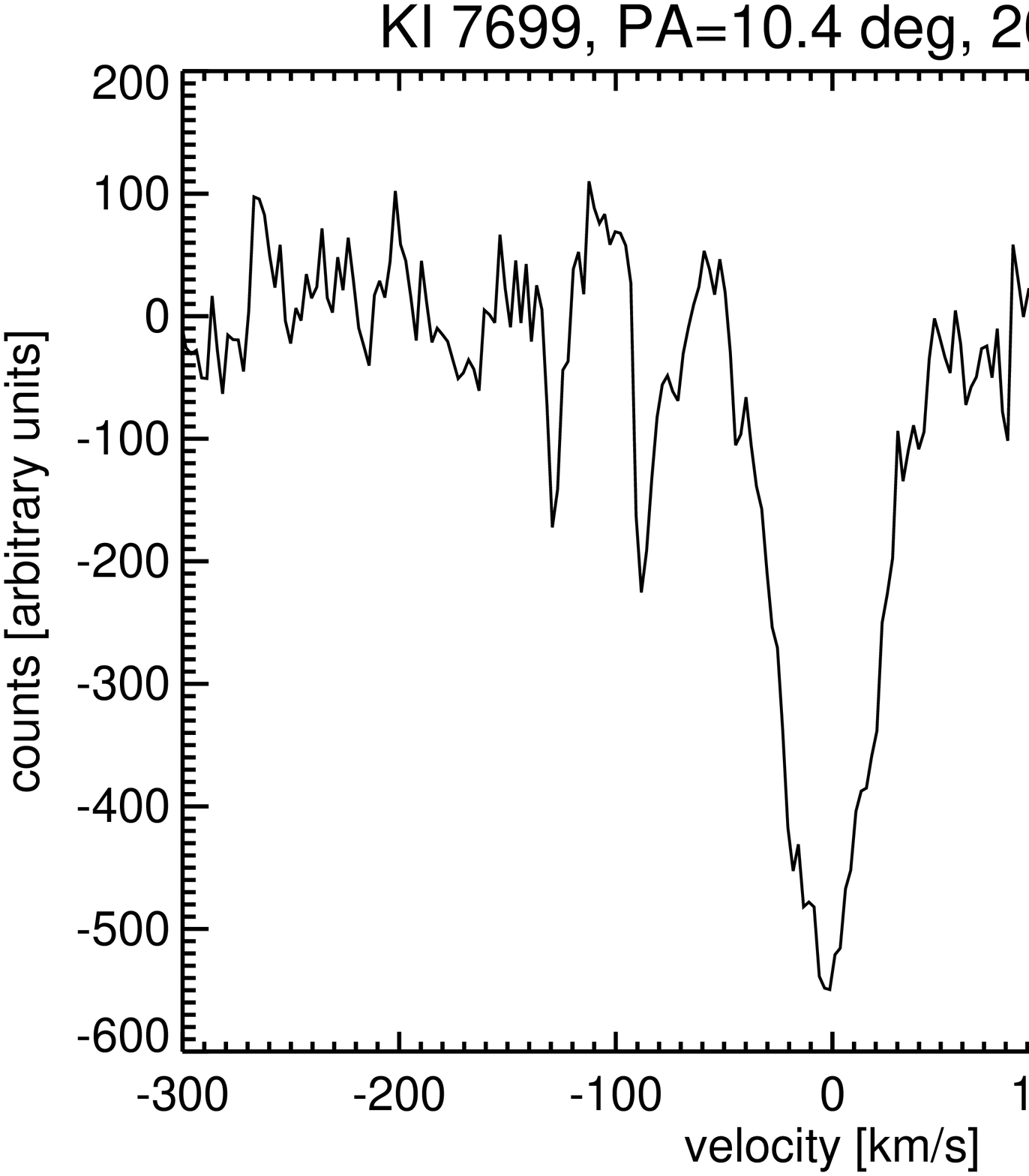}
 \hfill\includegraphics[width=5.7cm,clip]{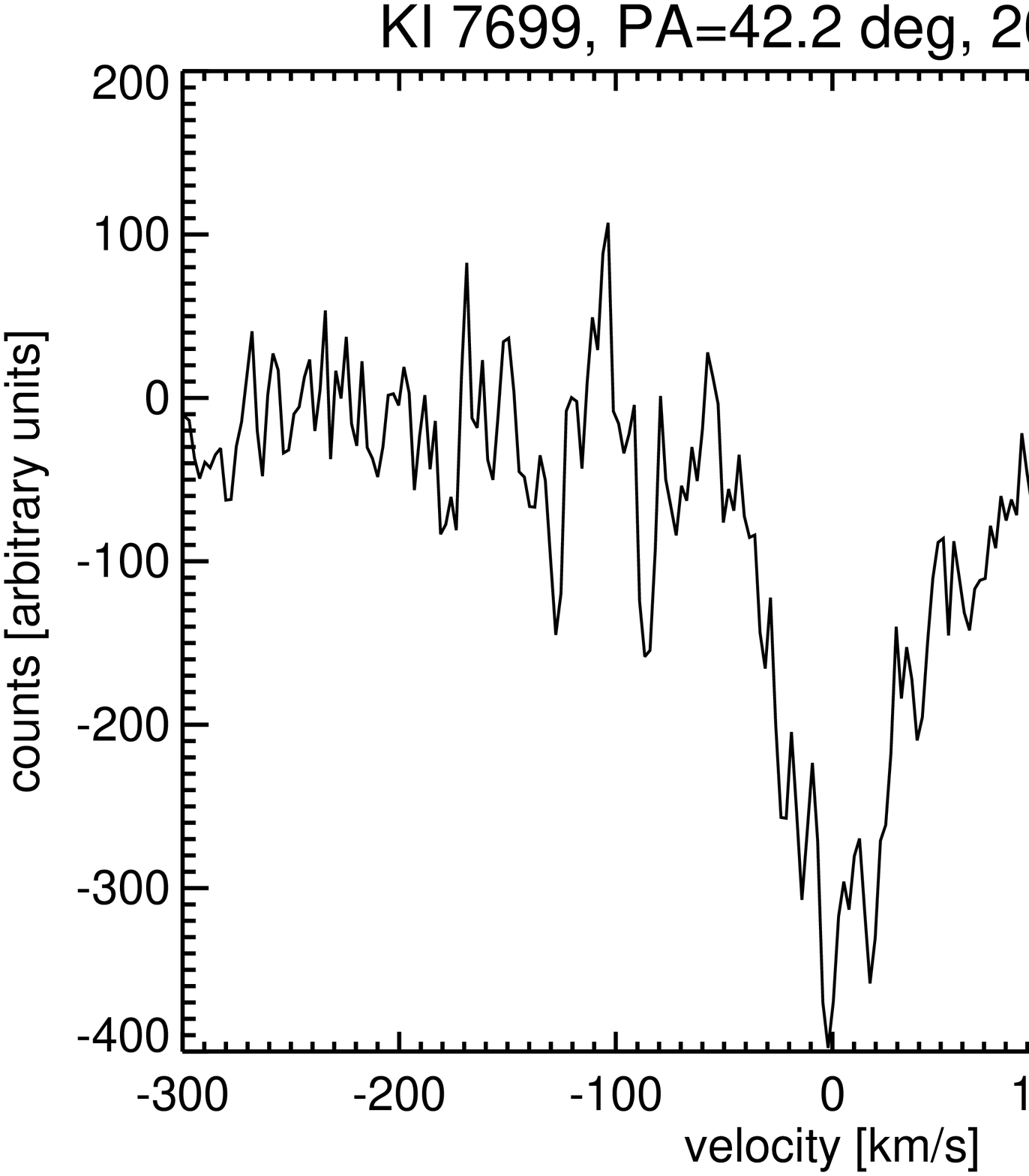}\hfill\mbox{}}
\hbox{\mbox{}
 \hfill\includegraphics[width=5.7cm,clip]{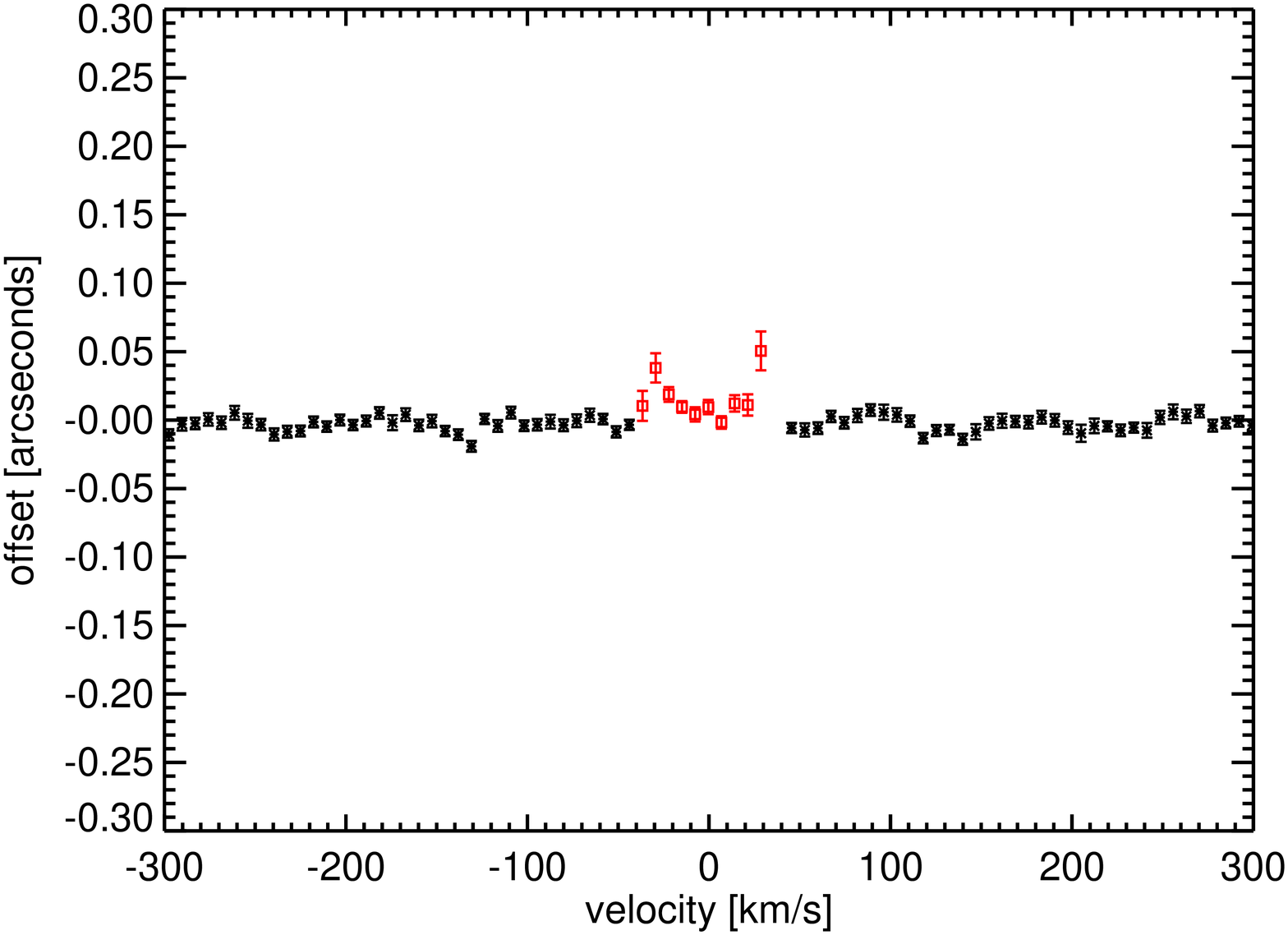}
 \hfill\includegraphics[width=5.7cm,clip]{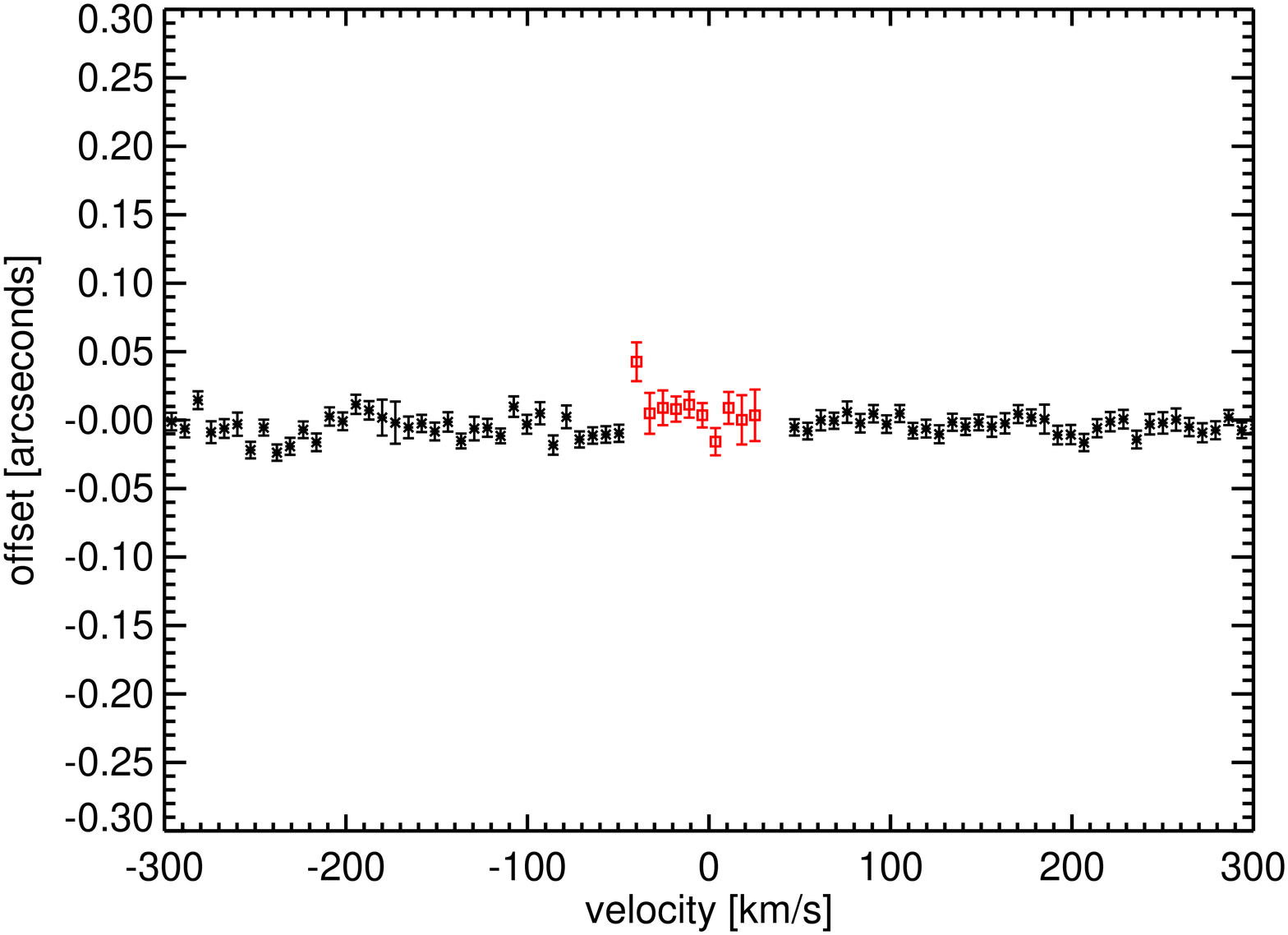}
 \hfill\includegraphics[width=5.7cm,clip]{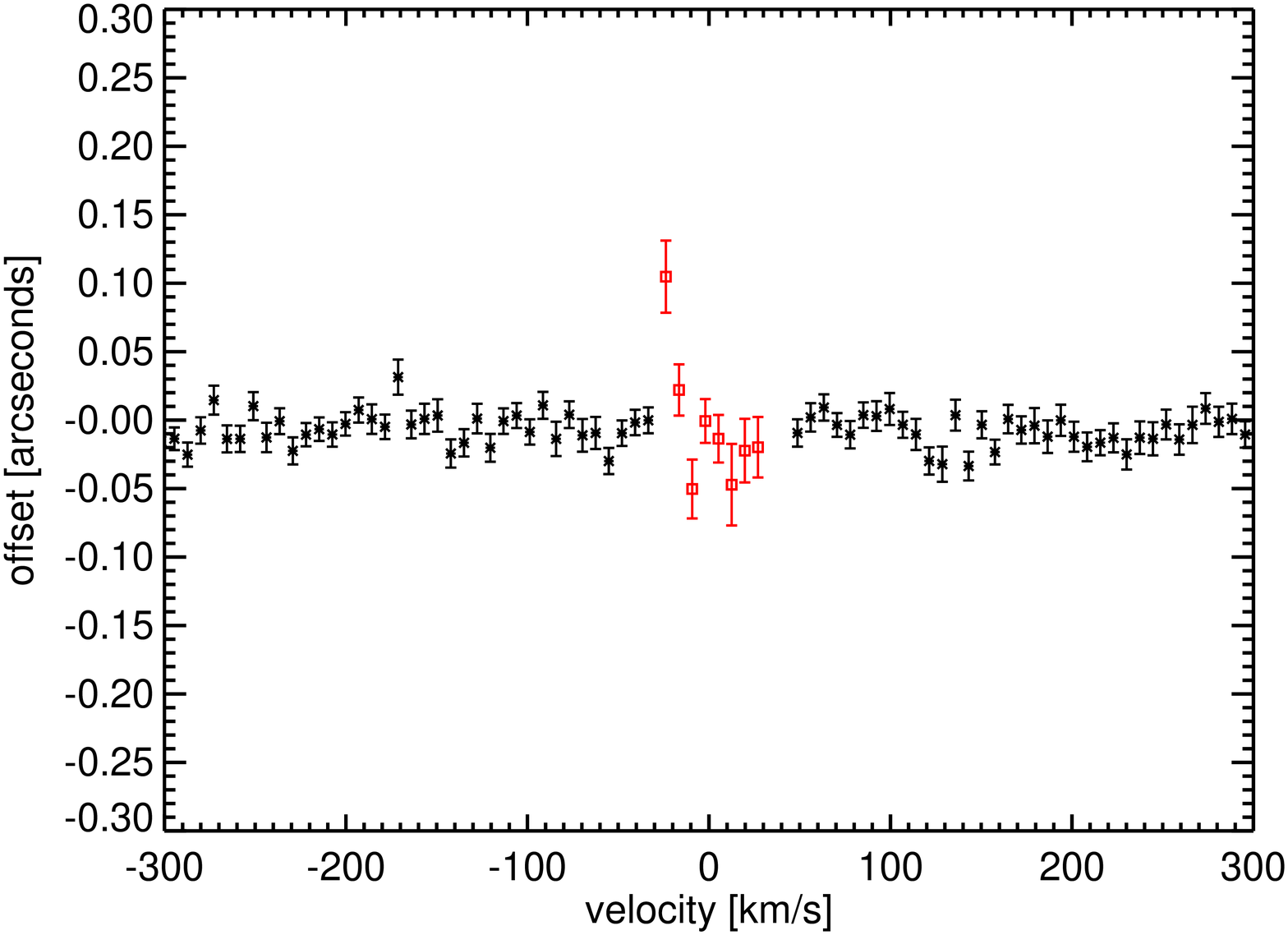}\hfill\mbox{}}
\hbox{\mbox{}
 \hfill\includegraphics[width=5.7cm,clip]{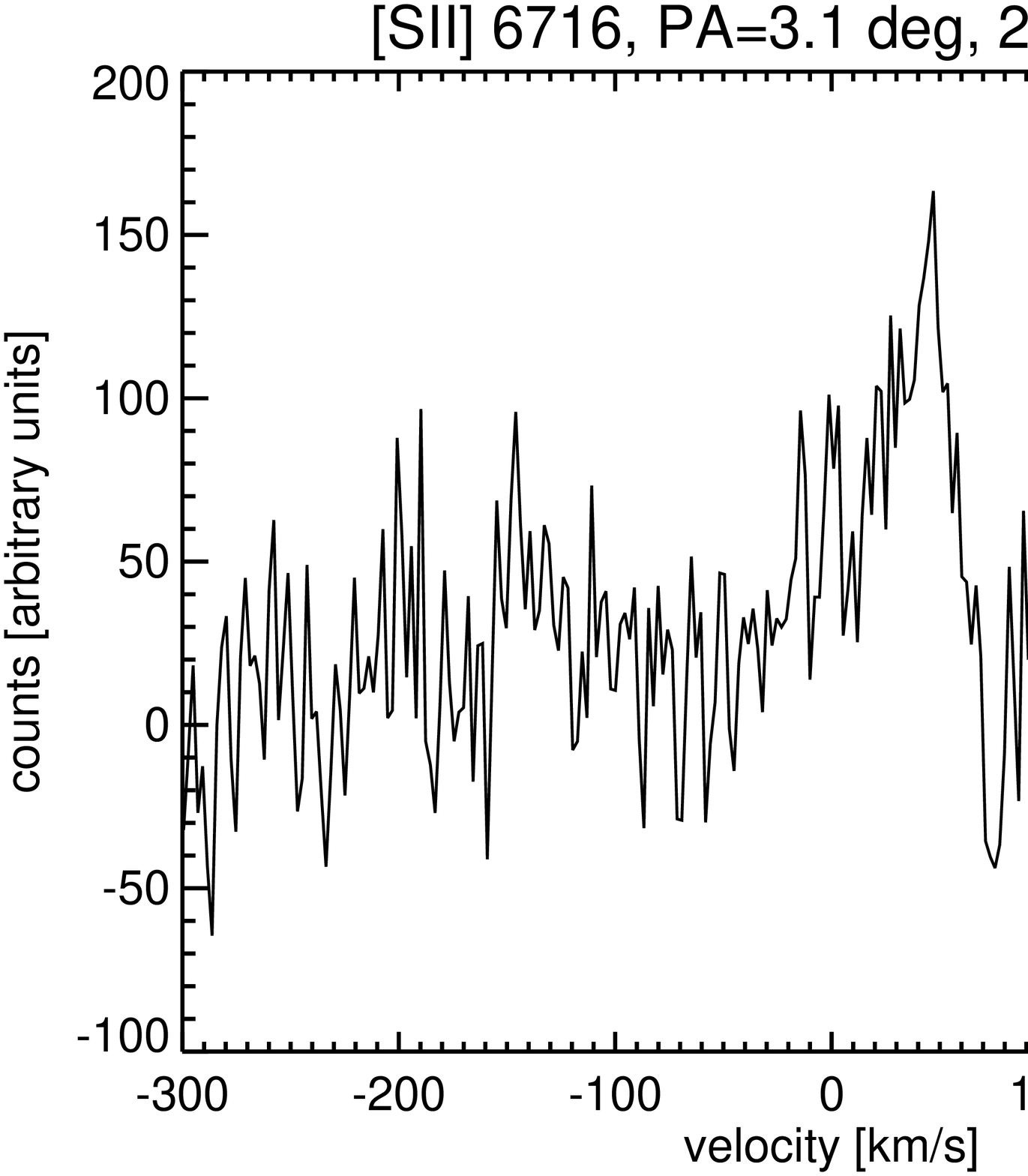}
 \hfill\includegraphics[width=5.7cm,clip]{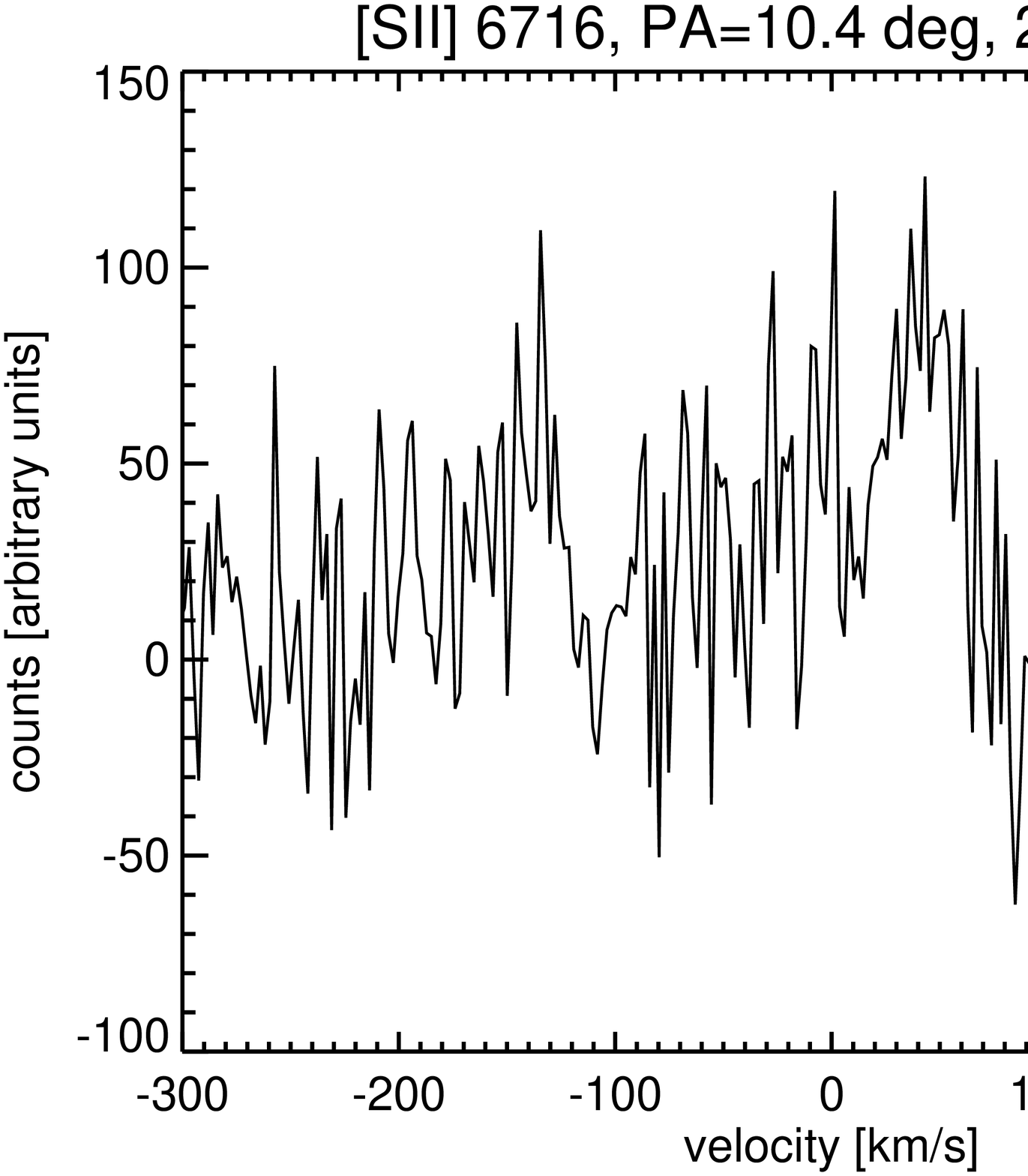}
 \hfill\includegraphics[width=5.7cm,clip]{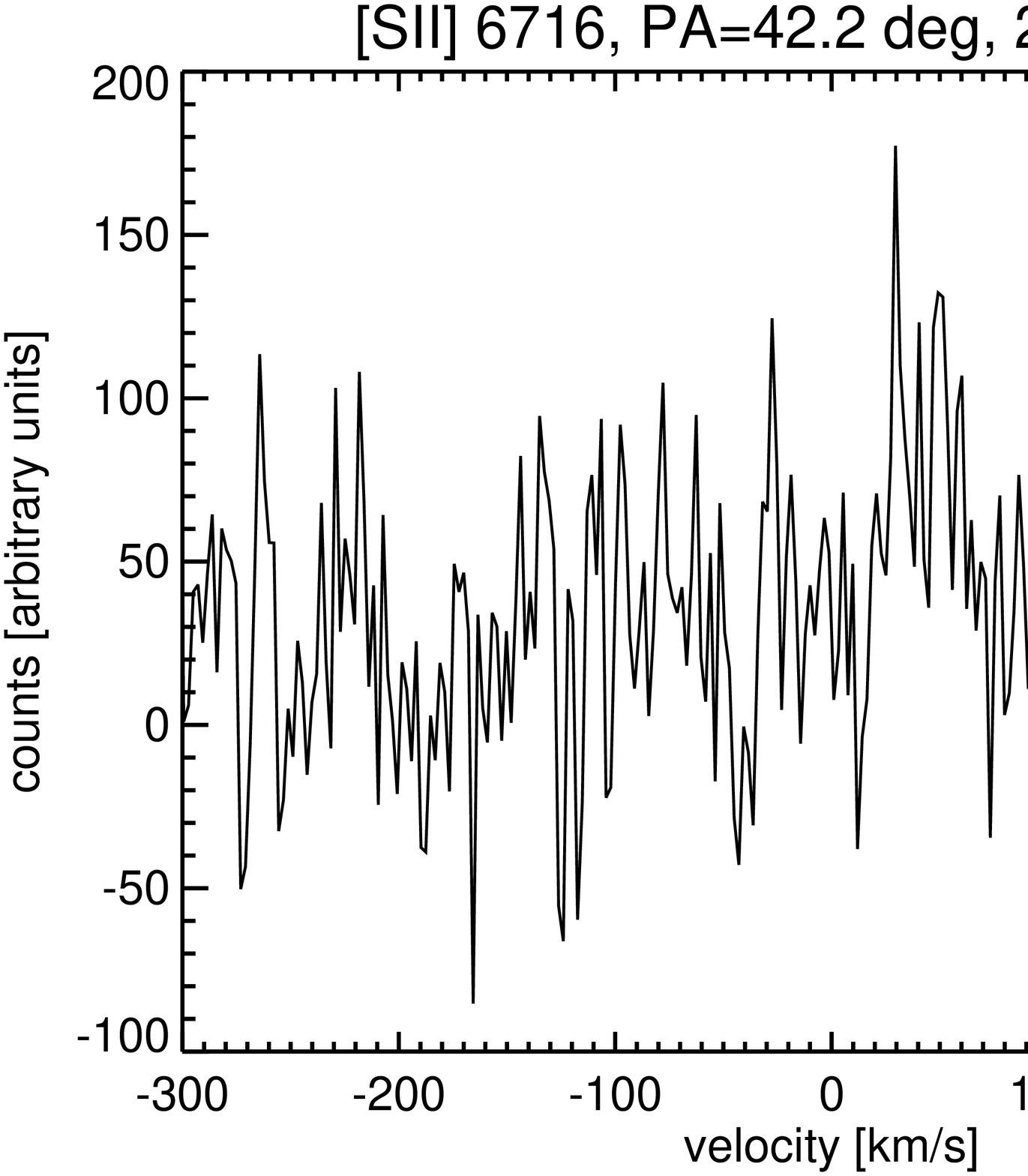}\hfill\mbox{}}
\hbox{\mbox{}
 \hfill\includegraphics[width=5.7cm,clip]{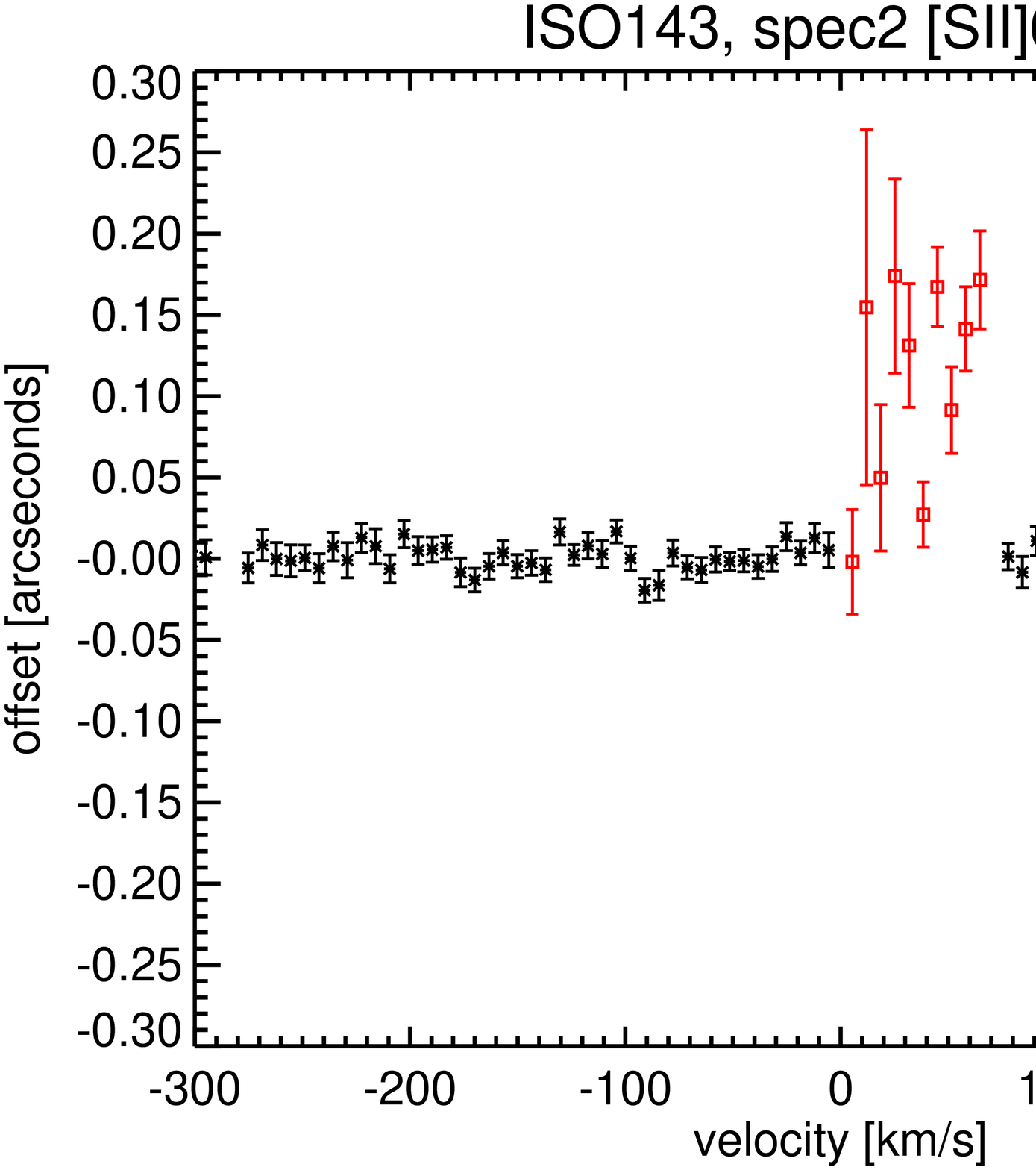}
 \hfill\includegraphics[width=5.7cm,clip]{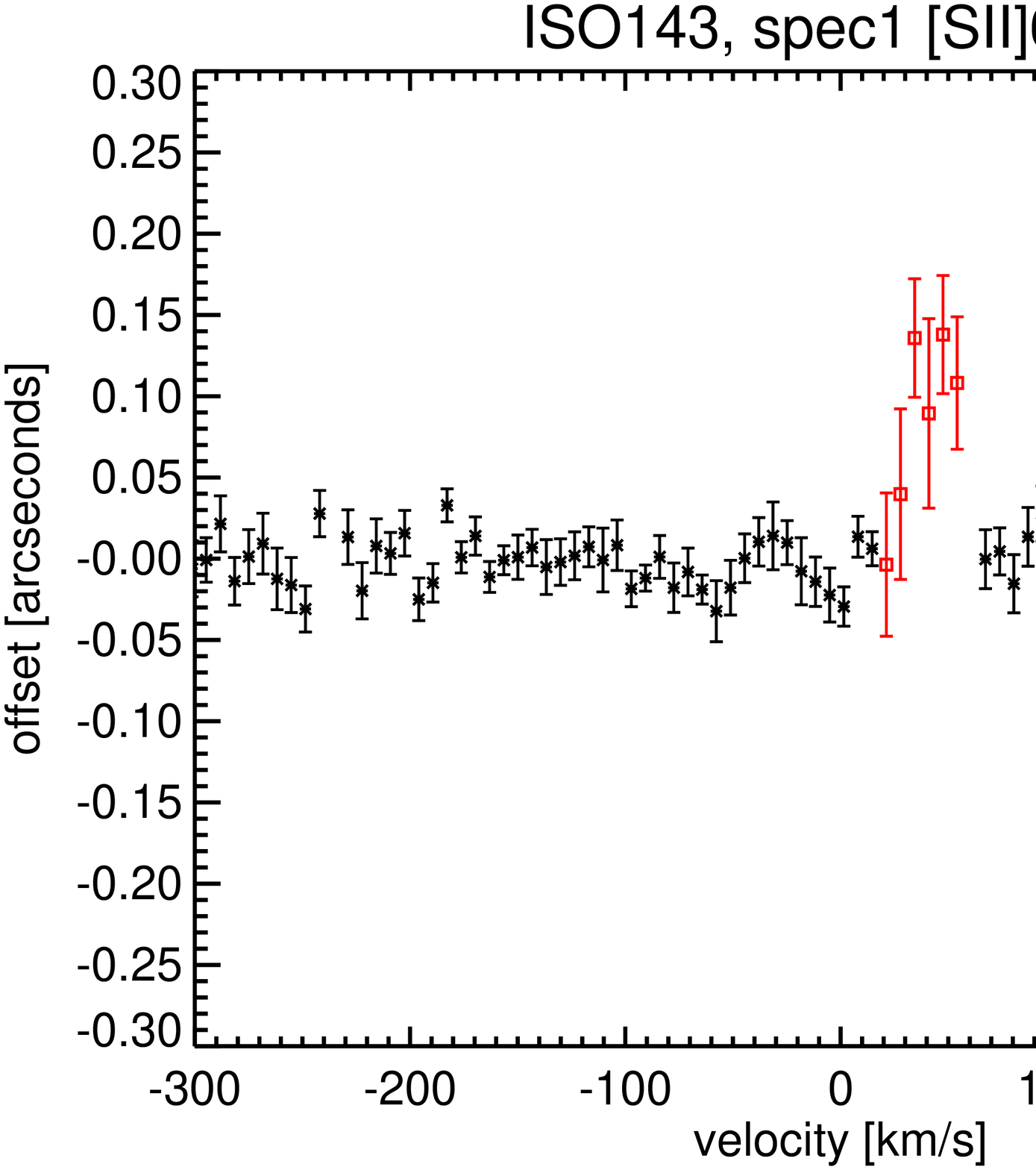}
 \hfill\includegraphics[width=5.7cm,clip]{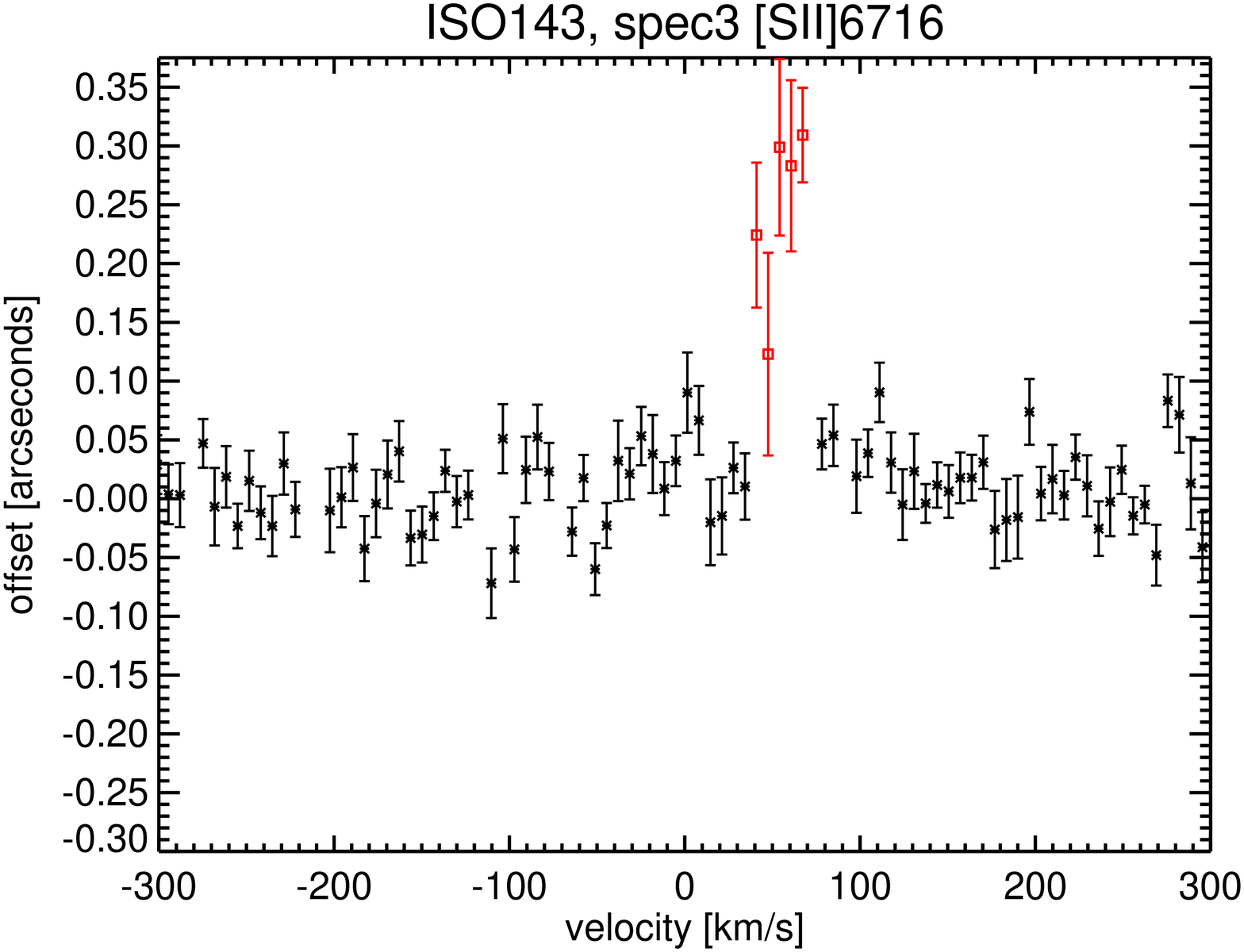}\hfill\mbox{}}
\hbox{\mbox{}
 \hfill\includegraphics[width=5.7cm,clip]{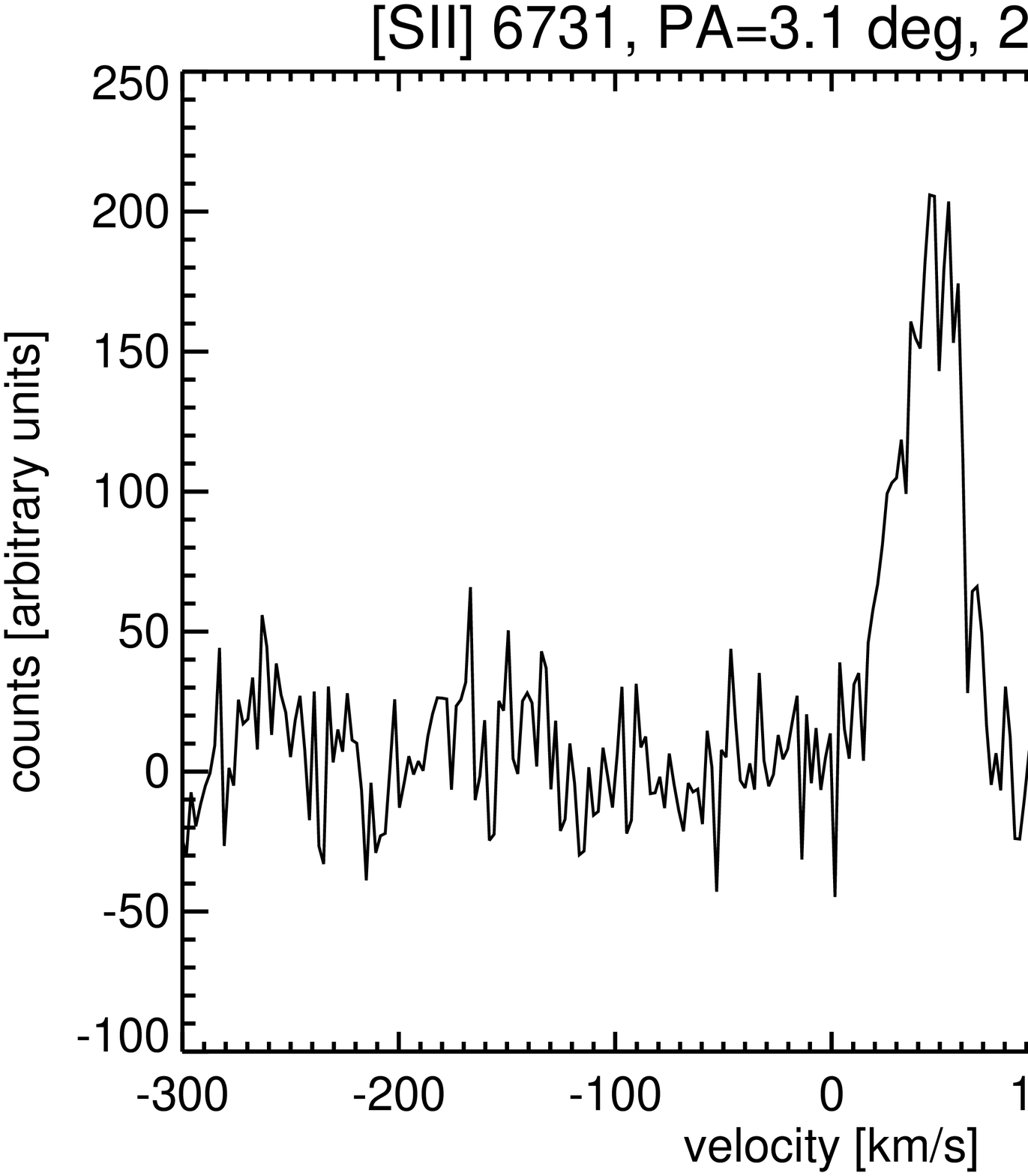}
 \hfill\includegraphics[width=5.7cm,clip]{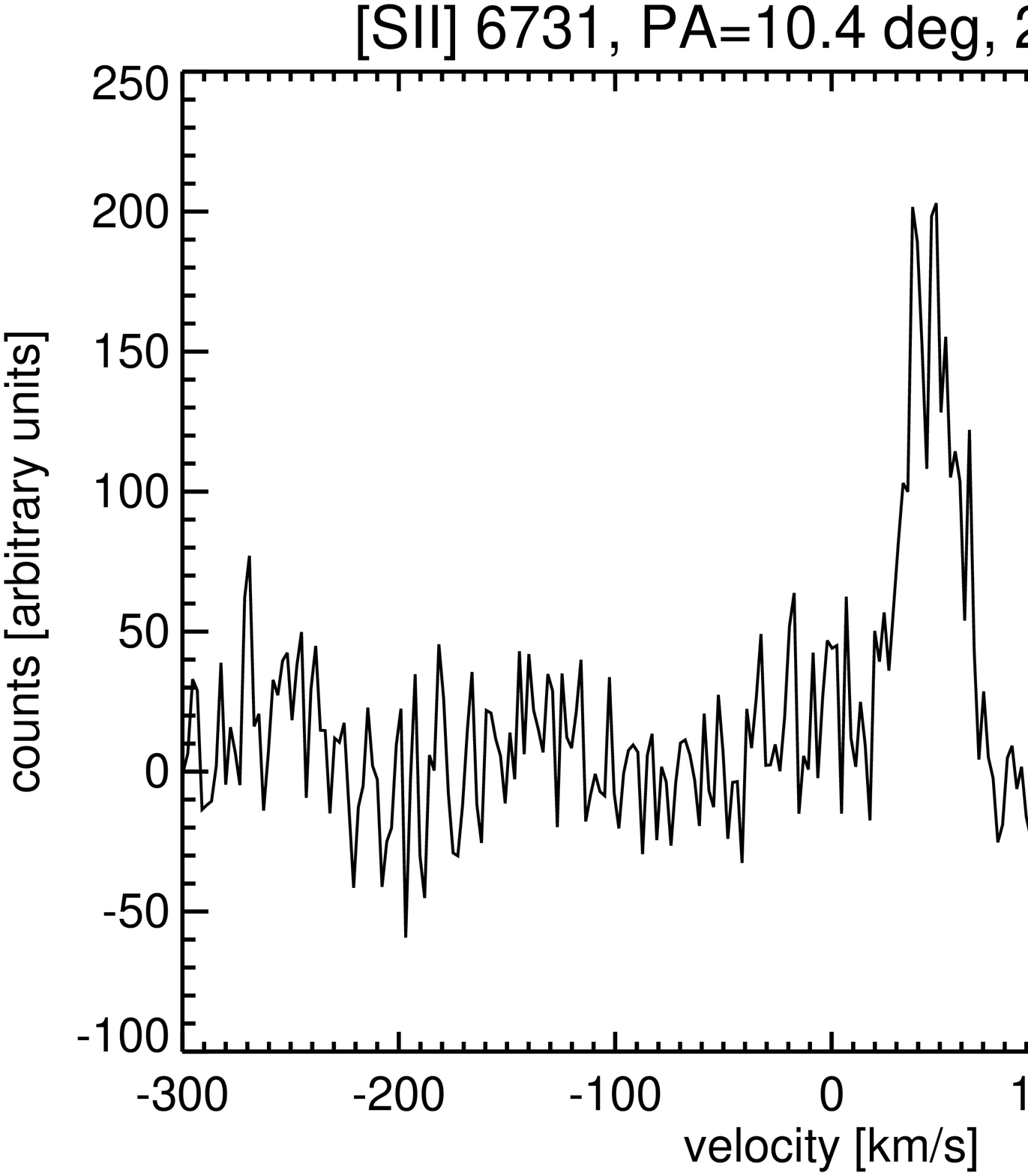}
 \hfill\includegraphics[width=5.7cm,clip]{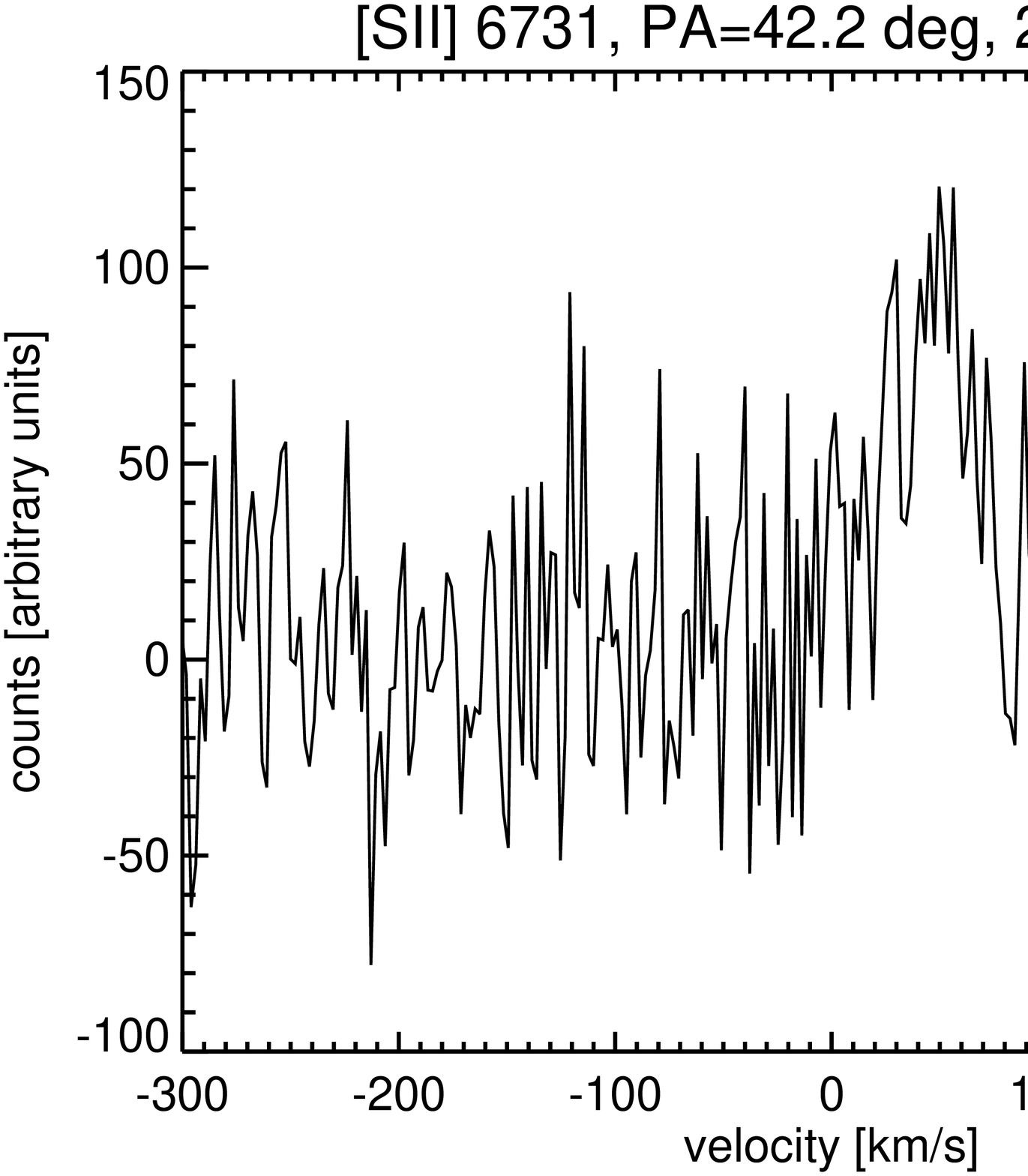}\hfill\mbox{}}
\hbox{\mbox{}
 \hfill\includegraphics[width=5.7cm,clip]{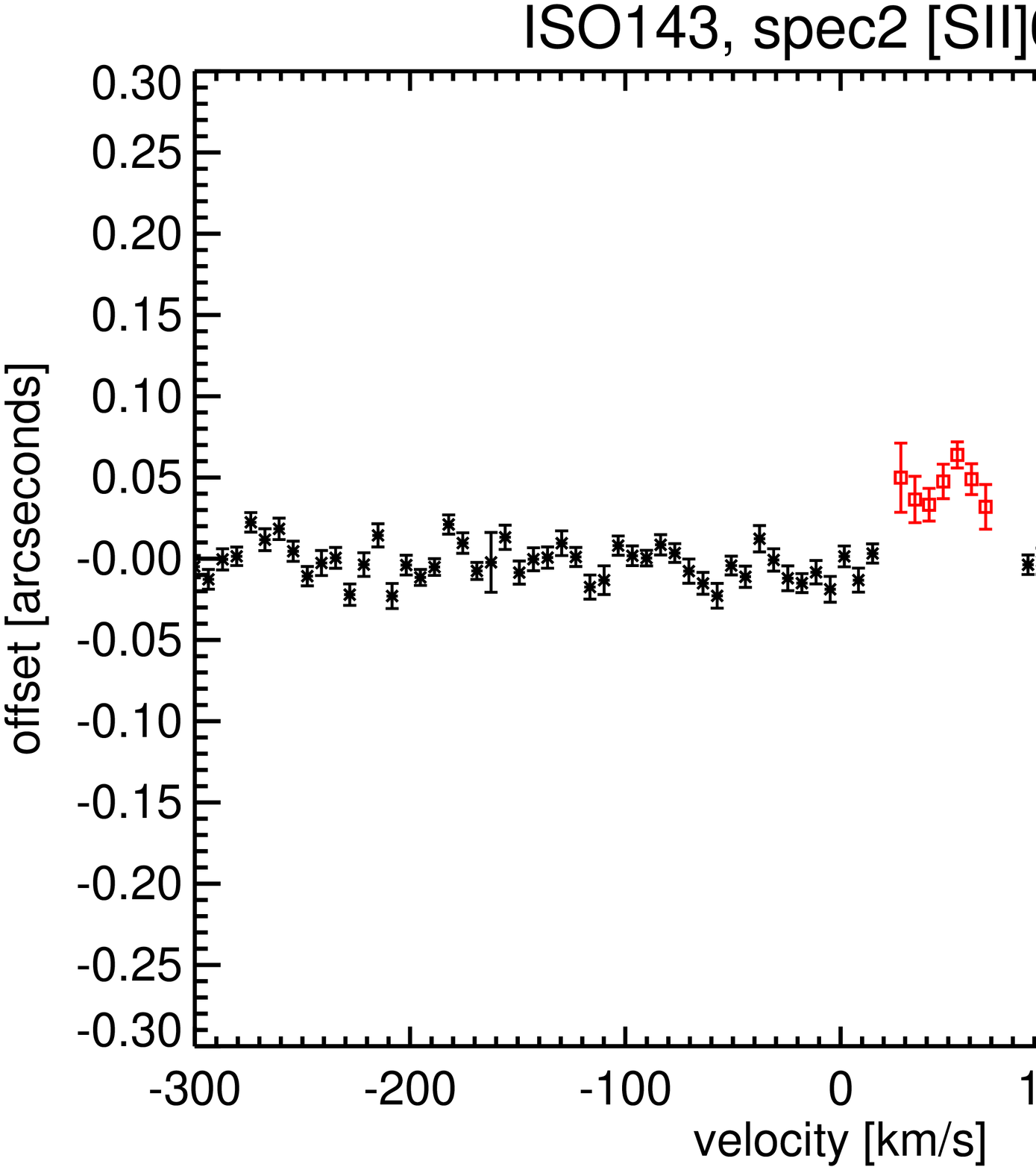}
 \hfill\includegraphics[width=5.7cm,clip]{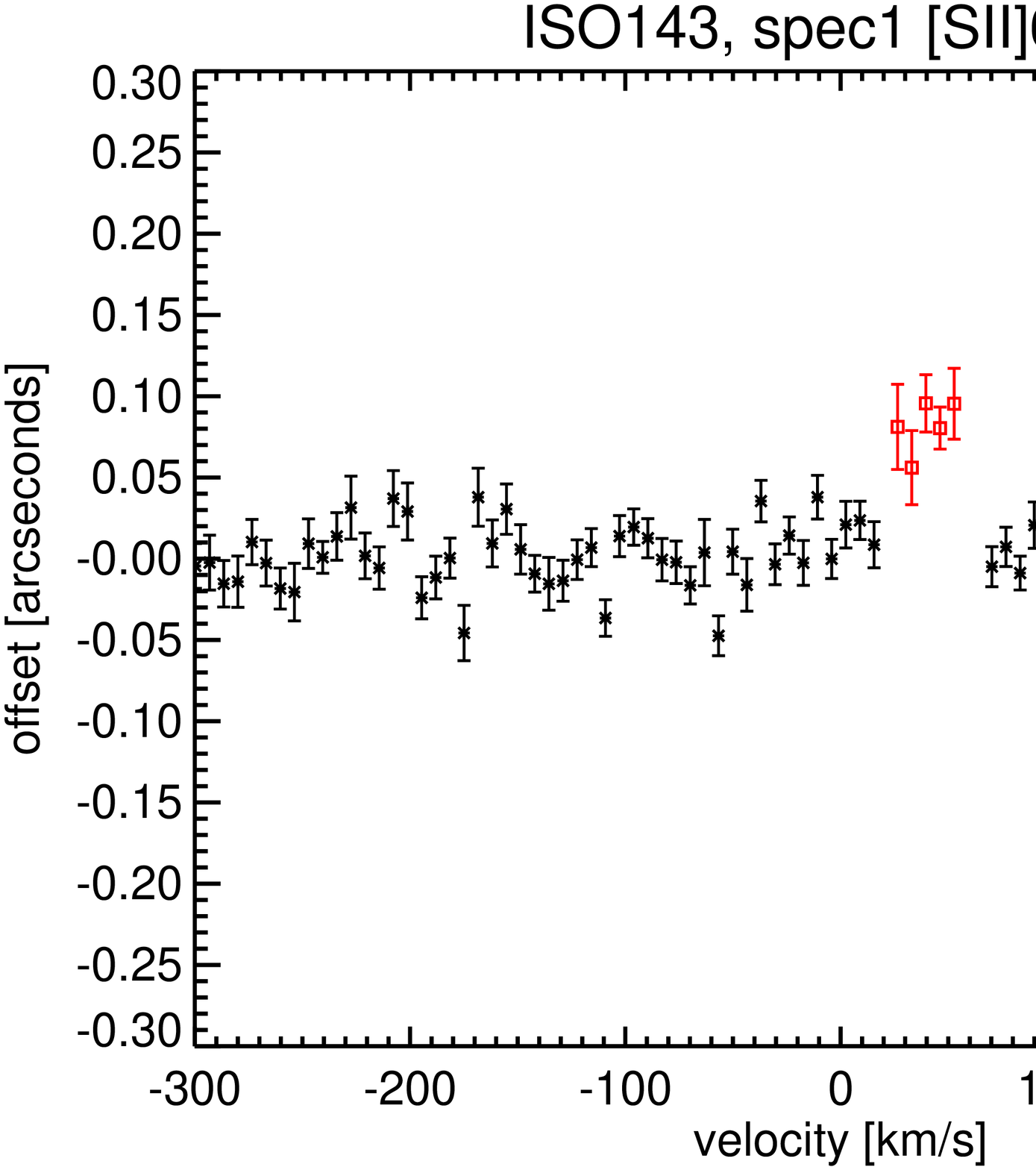}
 \hfill\includegraphics[width=5.7cm,clip]{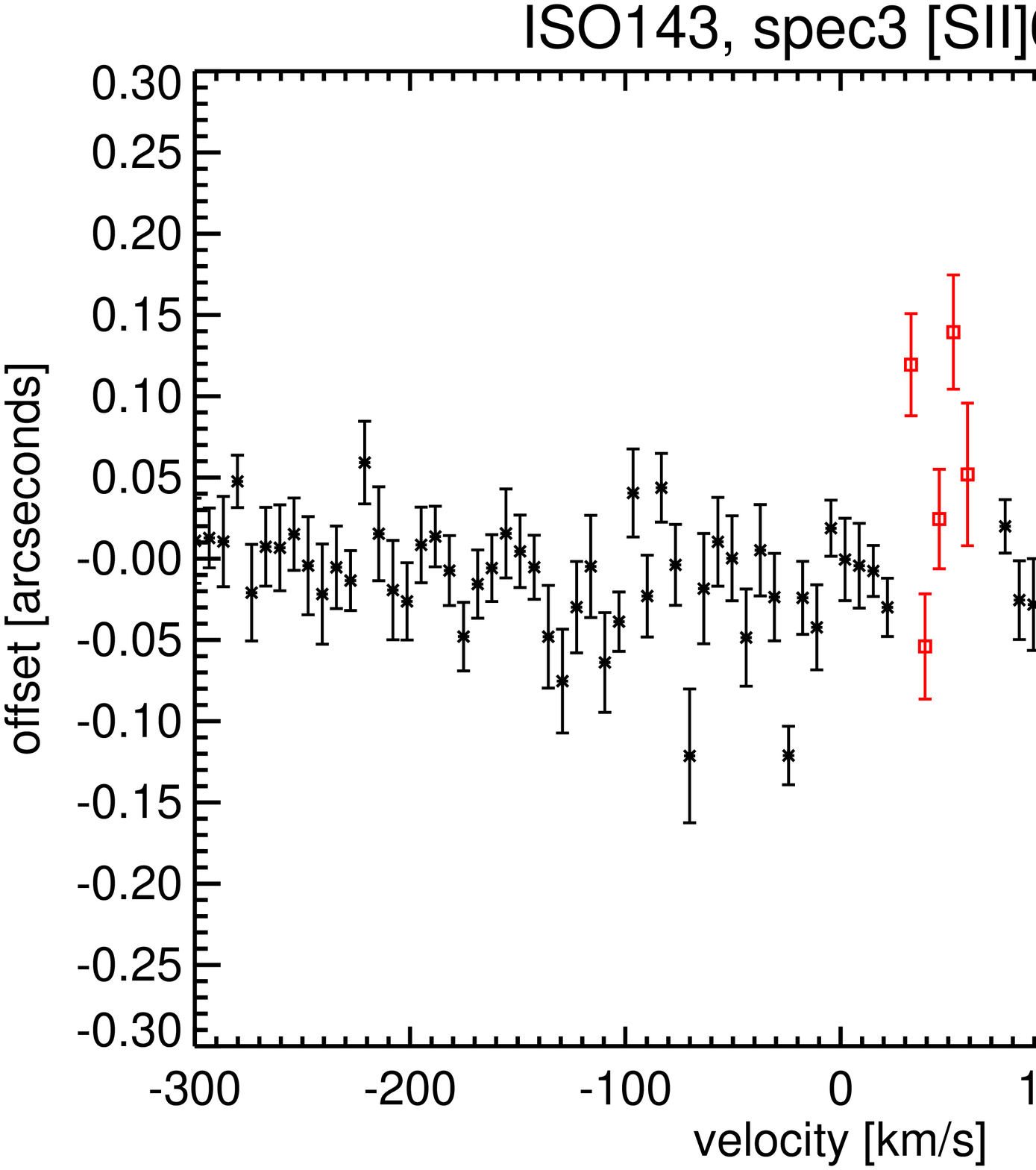}\hfill\mbox{}}
}
\caption{\label{fig:sa} 
Spectro-astrometric analysis of spectral lines of \iso143 for three different observing times and slit PAs. 
{\em Top row:} photospheric line K\,I\,$\lambda$7699
as a test for artifacts, {\em middle row:} FEL of [S\,II]\,$\lambda$6716,
{\em bottom row:} FEL of [S\,II]\,$\lambda$6731.
For each case, we display the 
line profiles in the top panels, which are the average of 8 pixel rows in the spatial direction centered on the continuum
after continuum subtraction, and the spectro-astrometric plots in the bottom panels, 
which show the spatial offset vs. radial velocity
of the continuum (black asterisks) and of the continuum subtracted spectral line (red squares).
Velocities are given relative to the stellar rest velocity $V_{0}$ of \iso143.
}
\end{figure*}

We used UVES spectra that were reduced, wavelength-, and flux-calibrated by means of the ESO UVES pipeline
in order to study activity related emission lines (Ca\,II, O\,I, He\,I, [S\,II], [Fe\,II]) regarding 
their line profile shapes (Sect.\,\ref{sect:lines}).
Furthermore, a spectro-astrometric analysis of the detected forbidden line
emission in [S\,II] was performed based on a custom-made reduction and wavelength calibration
procedure. 
After completing a standard CCD reduction of the raw data (bias and flatfield correction and
cosmic-ray elimination), a row-by-row wavelength calibration of the two-dimensional
spectra of individual echelle orders was done using the longslit package 
(fitcoords/transform) of IRAF. Finally, the sky was subtracted.
We measured the spectro-astrometric signature in
the resulting two-dimensional spectra by Gaussian fitting the spatial
profile at each wavelength of both the FELs and the 
adjacent continuum, following, e.g., Hirth et al. (1994a).
The spatial offset in the FEL was then computed relative to the continuum. 
To increase the signal-to-noise ratio (S/N), which is the limiting factor in applying 
spectro-astrometry to faint objects,
the data were binned in the wavelength direction. 
To rule out spectro-astrometric artifacts, which can be caused, for example, by an asymmetric PSF 
(Brannigan et al. 2006),
we applied spectro-astrometry to a photospheric absorption line (K\,I\,$\lambda$7699)
and demonstrated that the spectro-astrometric offset in this line does not exceed 50\,mas
(apart from one outlier, see Fig.\,\ref{fig:sa}, top row).
Further details on the spectro-astrometric analysis can be found in
Joergens et al. (2012).

\emph{Slit position angle.}
The UVES observations of \iso143 were obtained within the framework of a high-resolution
spectroscopic study of young brown dwarfs and VLMS that is
optimized for radial velocity work (e.g., Joergens 2008a). Therefore,
the slit orientation was kept aligned with the direction of the atmospheric 
dispersion, and, as a consequence, changed relative 
to the sky during the observations. 
Each 2$\times$1700\,sec exposure samples a range of on-sky slit position angles (PAs)
of about $\pm 8^{\circ}$, as given in Table\,\ref{tab:obslog}.
Despite the slit PA varying during the exposure,
the data nonetheless allowed the detection of outflows, 
as shown by our confirmation of the bipolar outflow of \iso217 (Joergens et al. 2012) 
and by the spectro-astrometry of \iso143 presented here.
The spectra of \iso143 were taken at mean slit PAs of 3$^{\circ}$, 10$^{\circ}$, and 40$^{\circ}$,
which enabled us to investigate the observed outflow extension as a function of the slit PA
(see Sect.\,\ref{sect:saresults}).

{\em Stellar rest velocity.}
A value of $V_{0}$=16.9\,km\,s$^{-1}$ was adopted for the stellar rest velocity of \iso143, 
which is the mean value of the radial velocities measured for \iso143 based on 
a cross correlation of many photospheric lines in the same spectra 
(Joergens 2012; cf. Table\,\ref{tab:obslog}).
All velocities in the following are given relative to this value of $V_{0}$.


\section{Results} 
\label{sect:results}

\subsection{Observed emission lines}
\label{sect:lines}

The observations of \iso143 at high spectral resolution with UVES reveal line emission in 
Ca\,II, O\,I, He\,I, [S\,II],  and [Fe\,II], 
which are signatures of chromospheric activity, ongoing accretion, winds, and/or outflowing material. 
We studied their line profile shapes, measured their equivalent widths (EWs) and full-width-at-half-maximum values (FWHMs)
in order to localize their formation sites in the system. 
For this purpose, we used the UVES spectra (Table\,\ref{tab:obslog})
after reduction, wavelength-, and flux-calibration by the ESO UVES pipeline. 
Furthermore, a polynomial fit to the continuum emission adjacent to each line was subtracted from the emission
line regions.
Several emission line profiles of \iso143 display complex shapes 
that obviously consist of more than one emission component.
We decomposed these line profiles into its constituent components using the IRAF task 
ngaussfit. We fitted up to three Gaussian functions to the profiles 
leaving the peak flux, velocity at the peak, and FWHM for each component as free parameters.
We furthermore measured EWs for each line component.
Table\,\ref{tab:ew} lists the observed emission lines, the determined peak radial velocities $V$, FWHMs, and EWs
for each individual component.
Details on the individual emission lines are presented in the following.

\begin{figure*}
\vbox{
\hbox{\mbox{}
 \hfill\includegraphics[width=5.7cm,clip]{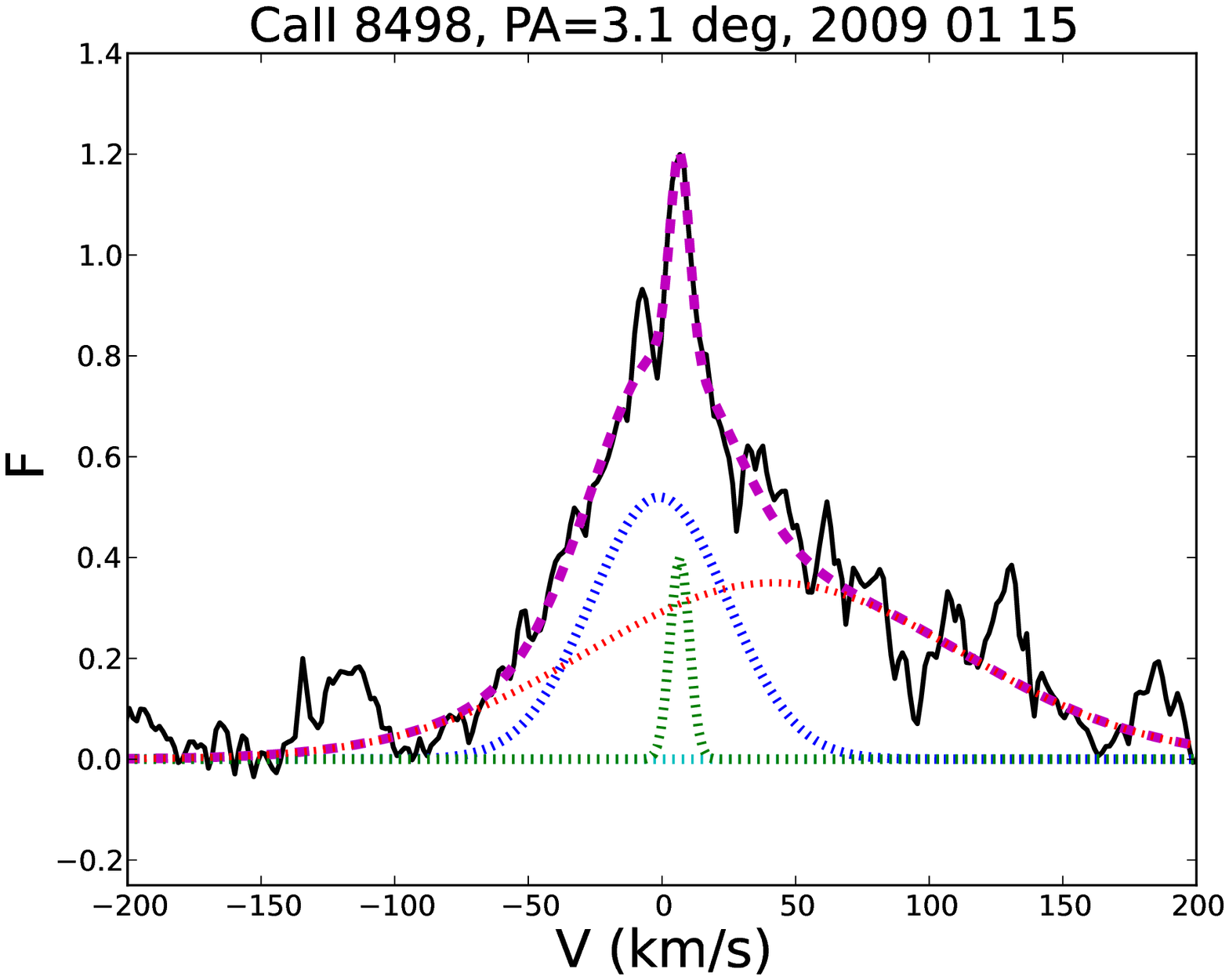}
 \hfill\includegraphics[width=5.7cm,clip]{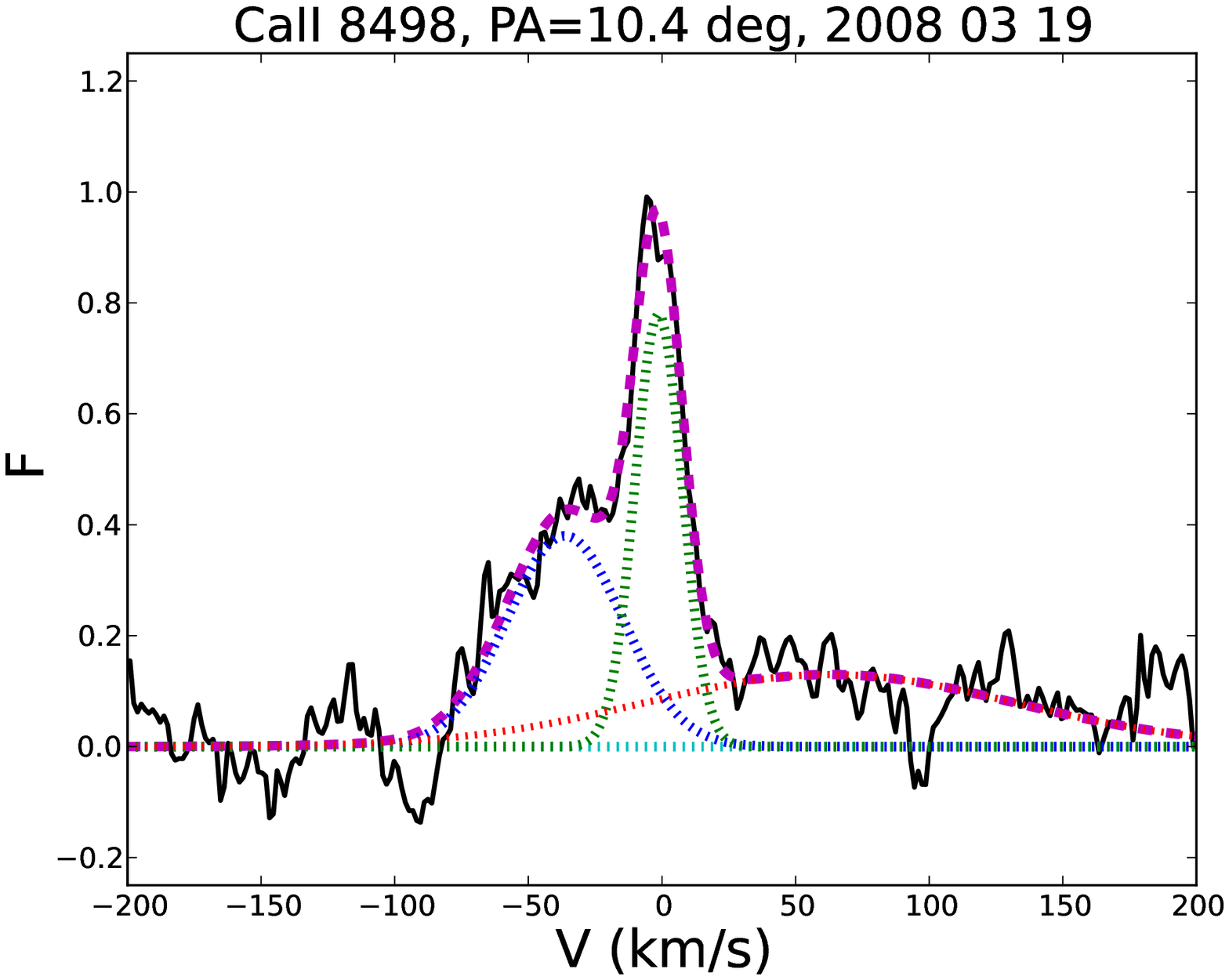}
 \hfill\includegraphics[width=5.7cm,clip]{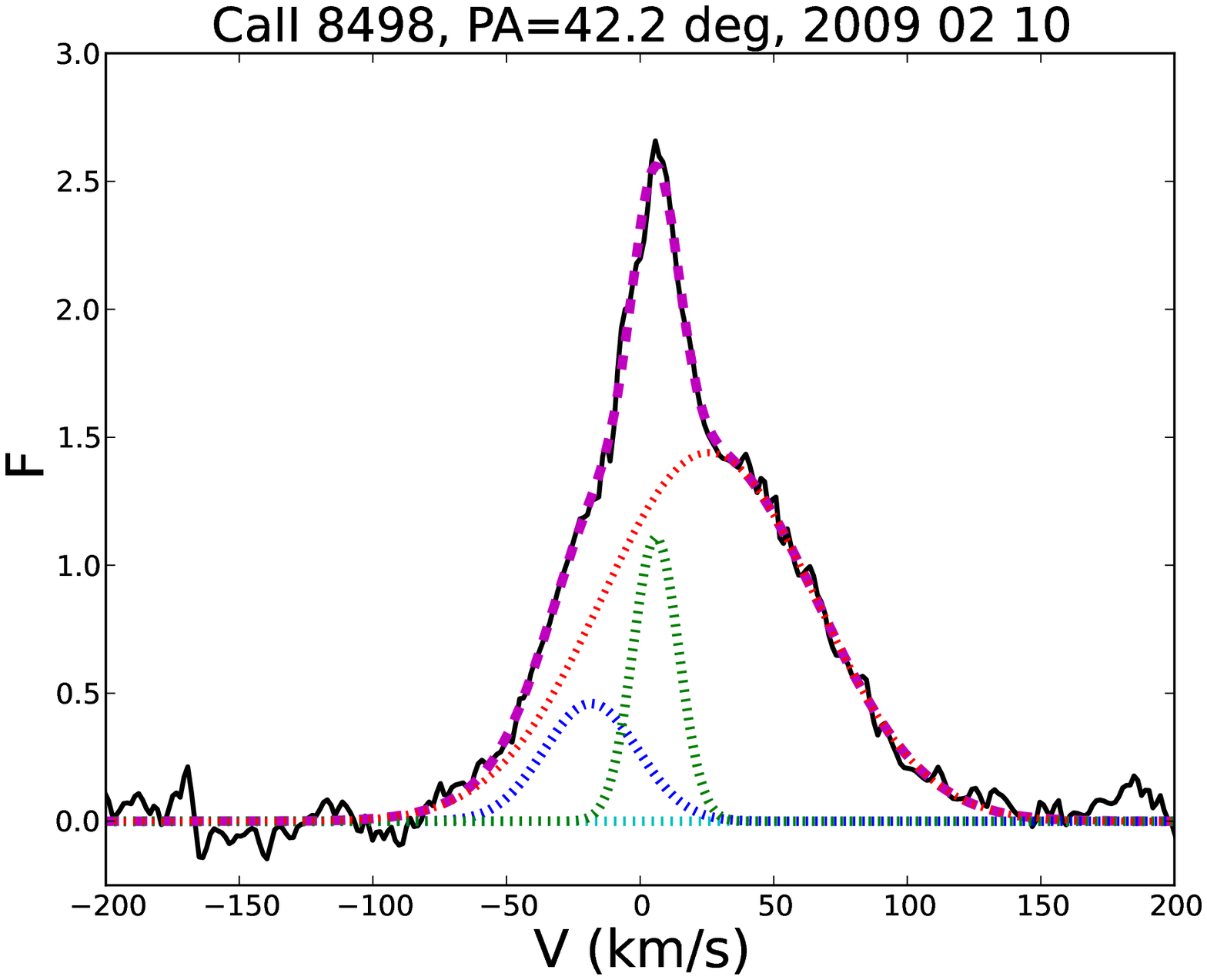}\hfill\mbox{}}
\hbox{\mbox{}
 \hfill\includegraphics[width=5.7cm,clip]{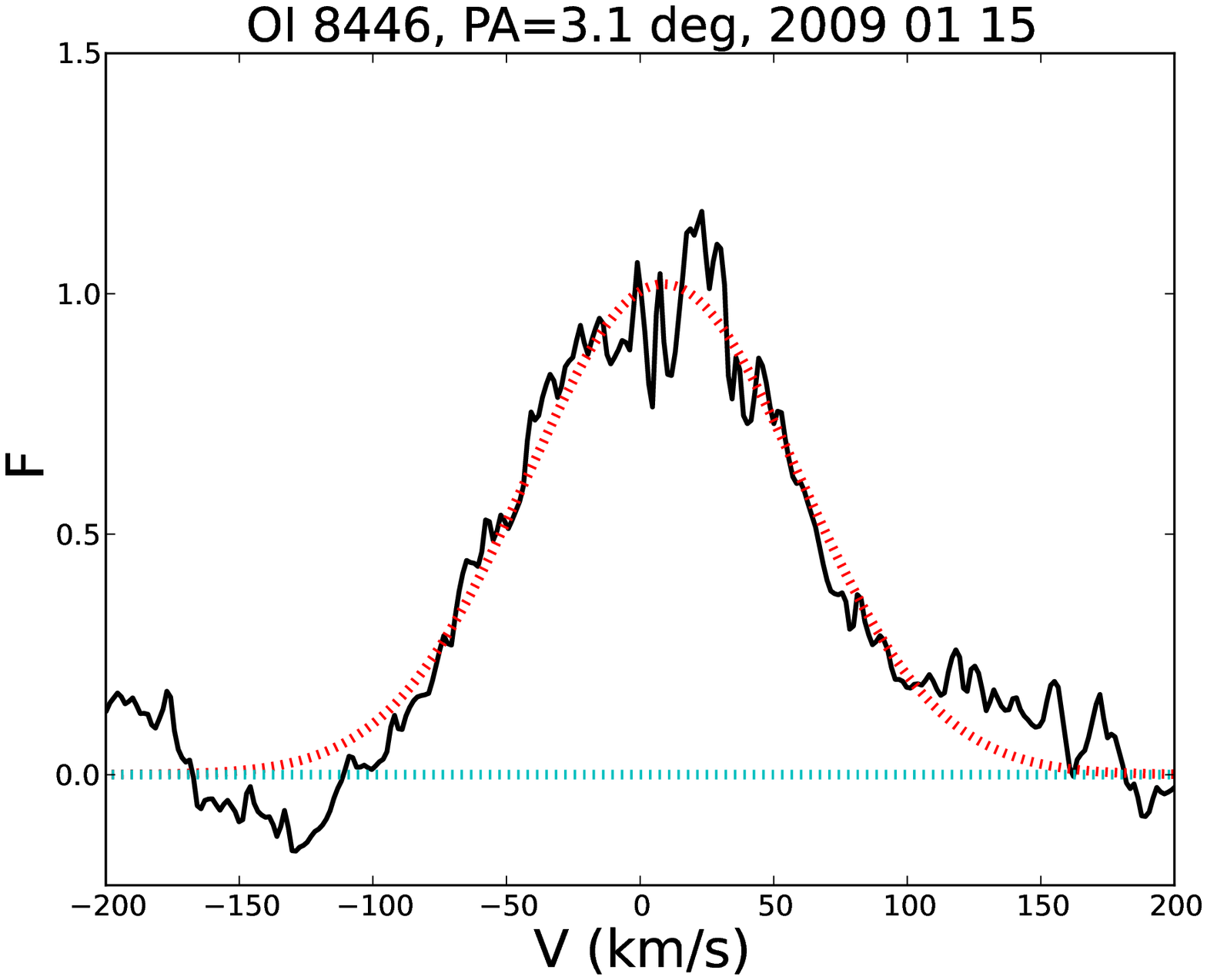}
 \hfill\includegraphics[width=5.7cm,clip]{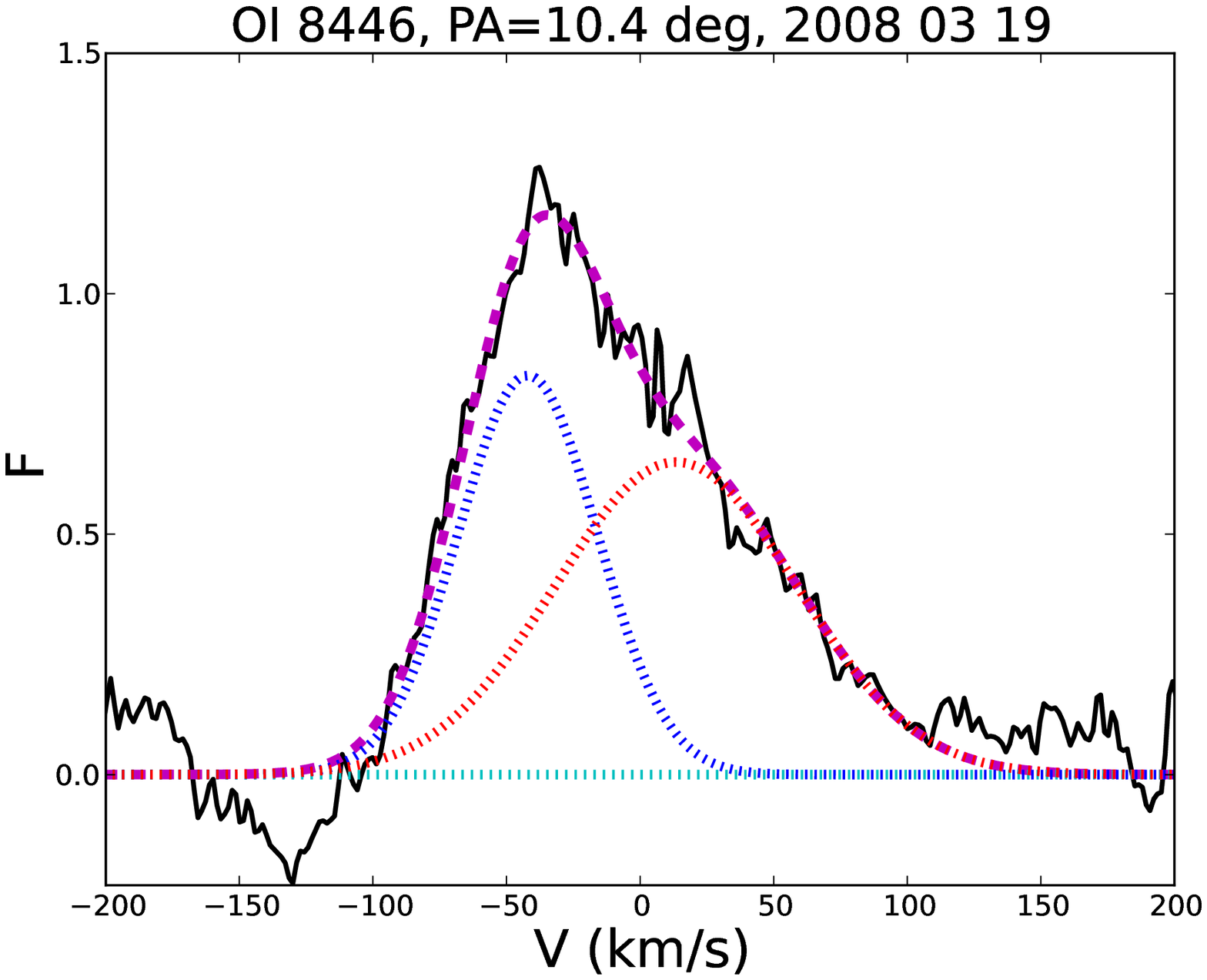}
 \hfill\includegraphics[width=5.7cm,clip]{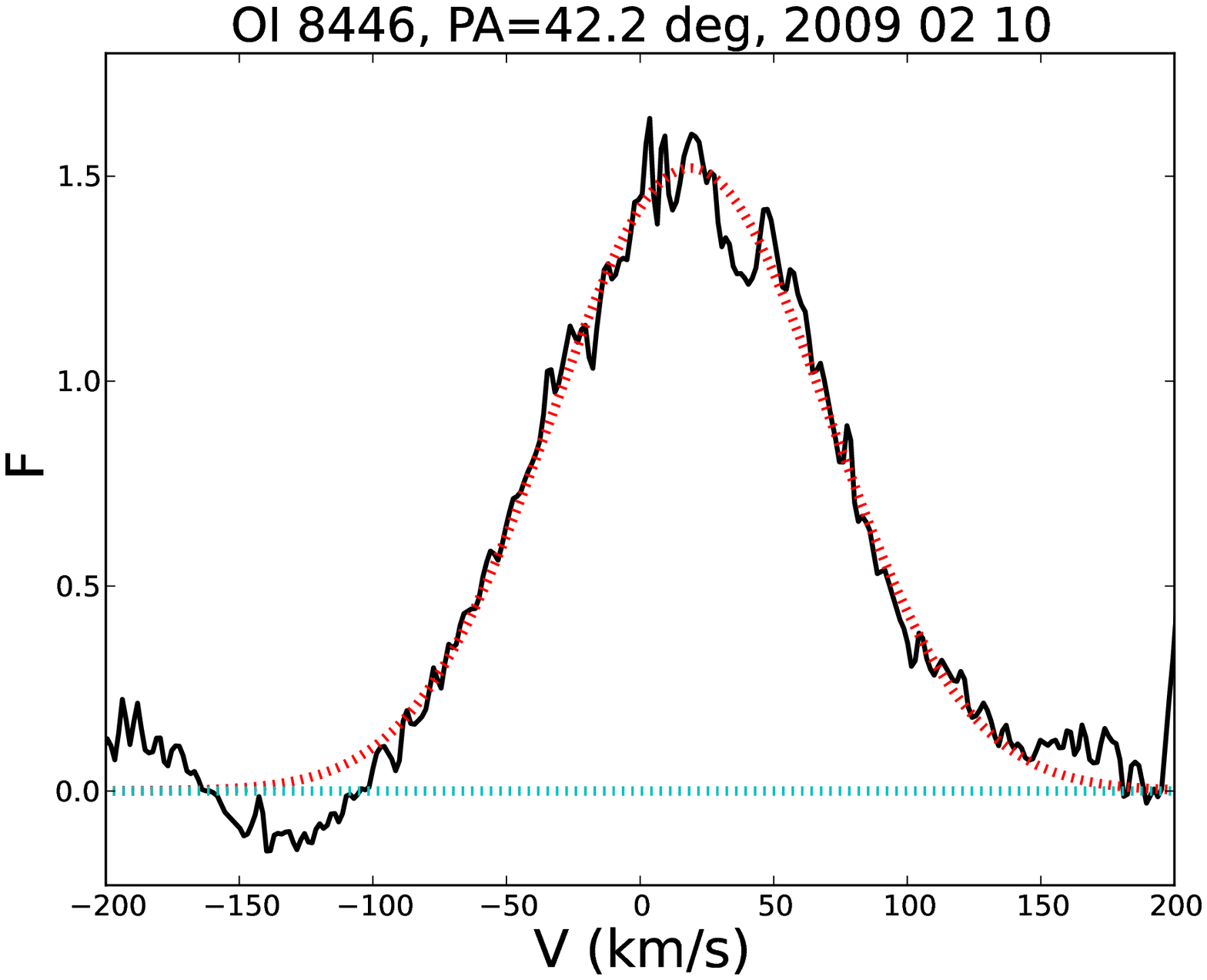}\hfill\mbox{}}
}
\caption{\label{fig:lp} 
Emission lines of Ca\,II\,$\lambda$8498 ({\em top row}) and O\,I\,$\lambda$8446 ({\em bottom row}) 
in UVES spectra of \iso143
at different observing times in Mar. 2008, Jan. 2009, and Feb. 2009 in the order of increasing slit PA 
after continuum subtraction. The flux is given in arbitrary units.
Gaussian fitting shows that the Ca\,II line can be decomposed into three components: 
(i) a narrow one centered close to zero velocity (probably chromospheric, green dotted line),
(ii) a broader blue-shifted one (possibly produced in a wind, blue dotted line), and
(iii) a very broad red-shifted one (possibly produced in magnetospheric infall, red dotted line).
The pink dashed line is the sum of the Gaussian functions.
The O\,I line can be fitted by a single red-shifted component in 
the two spectra from 2009 and by a blue- and a red-shifted component in
the spectrum from 2008 (the pink dashed line denotes here the sum of the two Gaussian functions). 
}
\end{figure*}

{\em Ca\,II IR emission.}
\iso143 exhibits a broad, asymmetric, and variable Ca\,II IR emission line at 8498\,{\AA}, which
clearly consists of more than one component. 
Figure\,\ref{fig:lp} displays this line at three different epochs
in 2008 and 2009 in the top panels. We note that the two other lines of this IR triplet 
are not covered by the observations because of a gap in wavelength between the 
CCD chips of the two-armed spectrograph.
A decomposition of the Ca\,II line profile into its constituent components 
demonstrates that it can be fitted by three Gaussian functions (Fig.\,\ref{fig:lp}): 
(i) a narrow emission component, which is centered close to zero velocity; 
(ii) a broader blue-shifted component; and 
(iii) a very broad red-shifted component.
The narrow emission component (peak at velocities between -1\,km\,s$^{-1}$ and 7\,km\,s$^{-1}$; 
FWHM of 9--21\,km\,s$^{-1}$, EW of 0.1--0.7\,${\AA}$)
is attributed to chromospheric activity giving its narrow profile 
centered close to the stars velocity. 
The variability of a few km\,s$^{-1}$ seen in this line 
component (Table\,\ref{tab:ew})
but not in the velocity of the star (Table\,\ref{tab:obslog}) could be explained by 
rotational modulation caused by inhomogenously distributed activity regions
in the chromosphere, such as plages 
(e.g., Huerta et al. 2008; Joergens 2008b; Prato et al. 2008 for 
T~Tauri related activity causing 
velocity variability of a few km\,s$^{-1}$; 
e.g., Berdyugina 2005 for Ca\,II emission produced in the chromosphere / chromospheric plages).  
The broader blue-shifted component in Ca\,II (peak at velocities
between -1\,km\,s$^{-1}$ and \mbox{-36\,km\,s$^{-1}$}; FWHM of 41--60\,km\,s$^{-1}$; EW of 0.6--0.9\,${\AA}$)
is suggested to be the signature of a wind that is expanding at a variable velocity.
The very broad red-shifted emission component (peak at
26--63\,km\,s$^{-1}$; FWHM of 94--165\,km\,s$^{-1}$; EW of 0.6--4.1\,${\AA}$),
which is significantly variable in both the peak velocity 
($\Delta V$ = 37\,km\,s$^{-1}$) and strength ($\Delta$ EW = 3.5\,${\AA}$),
is likely formed in the magnetospheric infall zone of the system
(cf. Muzerolle et al. 1998 for Ca\,II emission from infalling material in CTTS).
The sum of the three individual Gaussians 
(Fig.\,\ref{fig:lp}, top panels, pink dashed line), is the overall fit to the Ca\,II profile. 

{\em O\,I emission.}
\iso143 has strong, broad, and variable emission in the O\,I line at 8446\,{\AA} (Fig.\,\ref{fig:lp}, bottom panels).
This emission is centered on red-shifted velocities 
in the two spectra from 2009 and on blue-shifted velocities
in the spectrum from 2008.
We show that the line can be fitted by a single red-shifted Gaussian function for the two spectra in 2009
and that it can be decomposed into a blue- and a red-shifted component in the spectrum from 2008.
The profile of the red component of this O\,I line
has a very broad velocity distribution (FWHM of 106-121\,km\,s$^{-1}$)
and a peak at velocities that are moderately red-shifted (9-19\,km\,s$^{-1}$) but not as strong 
as seen in the red component of Ca\,II.  
The red component of O\,I is significantly variable in both 
the peak velocity 
($\Delta V$ = 10\,km\,s$^{-1}$) and strength ($\Delta$ EW = 3.4\,${\AA}$).
We attribute the emission in the red component mainly to disk accretion and, to a lesser extent, 
also to infalling material. 
The blue component that appears in this line in 2008 has approximately the same velocity (-42\,km\,s$^{-1}$) 
as the blue component in the Ca\,II line at the same epoch (-36\,km\,s$^{-1}$)
suggesting a common origin in a variable wind.

\begin{figure}[b]
\centering
\includegraphics[width=.85\linewidth,clip]{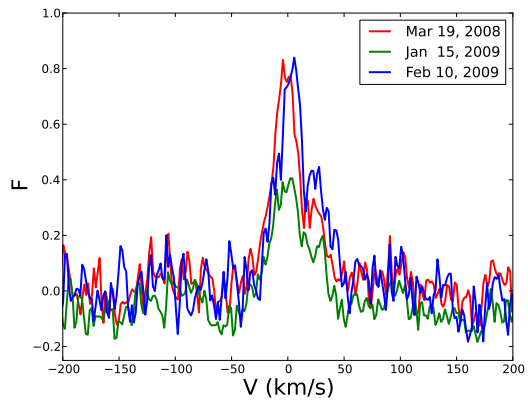}
\caption{
\label{fig:heI}
He\,I\,$\lambda$7065 emission line in UVES spectra of \iso143
at different observing times in 2008 and 2009 after continuum subtraction.
The flux is given in arbitrary units.
This He\,I line at 7065.2\,{\AA} is blended with another,
weaker He\,I line at 7065.7\,{\AA}, which 
shows up in the red line wing at a velocity of about +30\,km\,s$^{-1}$.
}
\end{figure}

\begin{figure}[b]
\centering
\includegraphics[width=.95\linewidth,clip]{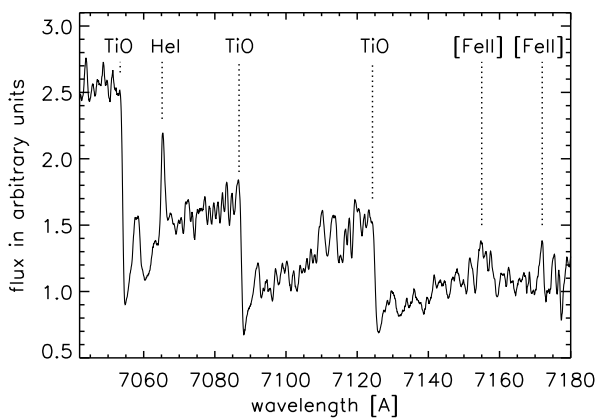}
\caption{
\label{fig:feII}
[Fe\,II]\,$\lambda\lambda$7155,7172 emission lines in a mean 
UVES spectrum of \iso143.
Also visible in this portion of the spectrum are TiO absorption bands with
band-heads at 7053\,{\AA}, 7087\,{\AA}, and 7124\,{\AA}, and the He\,I\,$\lambda$7065 emission line.
}
\end{figure}

{\em He\,I emission.}
We detect a prominent He\,I emission line at 7065\,{\AA} (Fig.\,\ref{fig:heI}-\ref{fig:feII}). 
The relatively narrow-band profile (FWHM of 26-35\,km\,s$^{-1}$) of this line
being centered on $V$=0 
and displaying little variability in its strength ($\Delta$ EW $\leq 0.1\,{\AA}$),
and the high 
temperature and density required to excite this transition 
indicates an origin in the chromosphere or the transition zone between the chromosphere and the corona 
(cf. e.g. Sicilia-Aguilar et al. 2012).
We note that a second weaker blended He\,I line at 7065.7\,{\AA} emerges
in the red wing of this He\,I line at 7065.2\,{\AA} (Fig.\,\ref{fig:heI}). 

{\em Forbidden [S\,II] emission.}
\iso143 shows strong emission in forbidden lines of sulfur [S\,II] at 6716\,{\AA} and 6731\,{\AA}
(Fig.\, \ref{fig:sa})
with a velocity at the line peak of 41-50\,km\,s$^{-1}$, a FWHM of 27-33\,km\,s$^{-1}$, 
and an EW of 0.2\,{\AA} for [S\,II]$\lambda$6716 and of 0.4--0.5\,{\AA} for [S\,II]$\lambda$6731
(Table\,\ref{tab:ew}).
The [S\,II] 
line profiles appear to vary little both in their strength ($\Delta$ EW $\leq 0.1\,{\AA}$)
and peak velocity ($\Delta V \lesssim$1\,km\,s$^{-1}$, with the exception of
the [S\,II]$\lambda$6716 profile on Feb. 2009; however, here the location of the peak was hampered by a low
S/N).
Spectro-astrometry of these lines 
(Sect.\,\ref{sect:saresults}) reveals that they are formed in an outflow at 
a distance from the central source of up to 30-50\,AU. 
For both [S\,II] lines, only a red-shifted component is visible. 
This implies that the outflow of \iso143 is intrinsically asymmetric and 
that there is either no or only little obscuration of the outflow seen in [S\,II] by the disk.

{\em Forbidden [Fe\,II] emission.}
We detect weak forbidden line emission of [Fe\,II] at 7155\,{\AA} and 
tentatively at 7172\,{\AA} (Fig.\,\ref{fig:feII}).
The peak of the [Fe II]\,$\lambda$7155 line seems to be around zero or low 
blue-shifted velocities; however, 
the poor S/N of this line hampers further quantitative analysis.
If confirmed, the observation of solely blue-shifted forbidden [Fe\,II] emission of 
a very low-mass outflow system that has a stronger red outflow component in [S\,II]
would resemble the case of the [Fe\,II] emission of the brown dwarf \iso217 (Joergens et al. 2012).

\begin{table*}
\begin{minipage}[t]{\columnwidth}
\begin{center}
\caption{
\label{tab:ew} 
Observed emission lines of \iso143.
}
\renewcommand{\footnoterule}{}  

\begin{tabular}{ccc|ccc|ccc|ccc}
		
\hline\hline  
\myrule
line & slit PA & date & \multicolumn{3}{c|}{blue component} & \multicolumn{3}{c|}{green component} & \multicolumn{3}{c}{red component}  \\[0.15cm]
     &         &      & $V$    & FWHM   & EW                & $V$    & FWHM   & EW                 & $V$    & FWHM   & EW               \\[0.15cm]
     & [deg]   &      & [km/s] & [km/s] & [\r{A}]           & [km/s] & [km/s] & [\r{A}]            & [km/s] & [km/s] & [\r{A}]          \\[0.15cm]
\hline 
\myrule
[S\,II]          & ~3.1 & 2009 01 15 & \ldots & \ldots & \ldots & \ldots & \ldots & \ldots & 40.8 &  27.3 & -0.2  \\
(6716.44\,{\AA}) & 10.4 & 2008 03 19 & \ldots & \ldots & \ldots & \ldots & \ldots & \ldots & 41.9 &  28.6 & -0.2  \\
                 & 42.2 & 2009 02 10 & \ldots & \ldots & \ldots & \ldots & \ldots & \ldots & 49.8 &  30.8 & -0.2  \\
\hline 
\myrule
[S\,II]          & ~3.1 & 2009 01 15  & \ldots & \ldots & \ldots & \ldots & \ldots & \ldots & 44.2 & 31.7 & -0.4 \\
(6730.82\,{\AA}) & 10.4 & 2008 03 19  & \ldots & \ldots & \ldots & \ldots & \ldots & \ldots & 44.5 & 26.7 & -0.5 \\
                 & 42.2 & 2009 02 10  & \ldots & \ldots & \ldots & \ldots & \ldots & \ldots & 44.6 & 32.5 & -0.4 \\
\hline 
\myrule
He\,I            & ~3.1 & 2009 01 15  & \ldots & \ldots & \ldots & -0.5 &  34.5 & -0.4 & \ldots & \ldots & \ldots \\
(7065.19\,{\AA}) & 10.4 & 2008 03 19  & \ldots & \ldots & \ldots & -2.3 &  26.2 & -0.5 & \ldots & \ldots & \ldots \\
                 & 42.2 & 2009 02 10  & \ldots & \ldots & \ldots &  2.4 &  28.1 & -0.5 & \ldots & \ldots & \ldots \\
\hline 
\myrule
O\,I             & ~3.1 & 2009 01 15  & \ldots & \ldots & \ldots & \ldots & \ldots & \ldots &  8.8 & 120.2 & -3.7  \\
(8446.36\,{\AA}) & 10.4 & 2008 03 19  &  -42.1 & 61.3   & -1.5   & \ldots & \ldots & \ldots & 13.5 & 106.1 & -2.1  \\
                 & 42.2 & 2009 02 10  & \ldots & \ldots & \ldots & \ldots & \ldots & \ldots & 18.8 & 121.2 & -5.5  \\
\hline 
\myrule
 Ca\,II          & ~3.1 & 2009 01 15  &  -1.0 & 60.0 & -0.9 &  6.6 &  8.5 & -0.1 & 42.0 & 164.8 & -1.7  \\
(8498.02\,{\AA}) & 10.4 & 2008 03 19  & -36.3 & 51.1 & -0.6 & -1.3 & 20.9 & -0.5 & 62.8 & 163.6 & -0.6  \\
                 & 42.2 & 2009 02 10  & -18.4 & 41.4 & -0.6 &  5.9 & 20.2 & -0.7 & 25.7 &  94.1 & -4.1  \\
\hline
\end{tabular}
\tablefoot{The listed entries are:
line with laboratory wavelength, slit PA, observing date, and for each individual component 
the velocity at the line peak, the FWHM, and EW. 
The laboratory wavelengths are taken from the 
NIST database (http://physics.nist.gov/asd3, Ralchenko et al. 2011).
The epochs are listed in the order of increasing slit PA, as 
in Table\,\ref{tab:fel}, for clarity.}
\end{center}
\end{minipage}
\end{table*}

\subsection{Results of spectro-astrometry}
\label{sect:saresults}

\begin{table*}
\begin{minipage}[t]{\columnwidth}
\centering
\caption{
\label{tab:fel} 
Outflow parameters from forbidden [S\,II] lines and mass accretion rate from the OI\,8446 line.
}
\renewcommand{\footnoterule}{}  
\begin{tabular}{cc|ccc|ccccc}
\hline
\hline
\myrule
Slit PA & Date & Line & Offset & V$_r$          & $n_{\rm{e}}$ & $L_{6731}$ & $\dot{M}_{out}$ & $L_{8446}$ & $\dot{M}_{acc}$ \\
$[$deg] &      &      & [mas]  & [km\,s$^{-1}$] & [cm$^{-3}$]  & $\lsun$    & [$\mperyr$]     & $\lsun$    & [$\mperyr$] \\
\hline
\myrule
\multirow{2}*{~3.1 } & \multirow{2}*{2009 01 15} & [S\,II]\,$\lambda$6716 & 172~($\pm$08) & 65 & \multirow{2}*{5600} & \multirow{2}*{6.8$\times$10$^{-7}$} & \multirow{2}*{3.1$\times$10$^{-10}$} & \multirow{2}*{3.6$\times$10$^{-5}$} & \multirow{2}*{1.8$\times$10$^{-8}$}\\
                 &            & [S\,II]\,$\lambda$6731 & ~64~($\pm$14) & 54 \\ 
                 &            &                                        &    & & & \\  
\multirow{2}*{10.4 } & \multirow{2}*{2008 03 19} & [S\,II]\,$\lambda$6716 & 138~($\pm$17) & 48 & \multirow{2}*{$> 2 \times 10^{4}$} & \multirow{2}*{8.5$\times$10$^{-7}$} & \multirow{2}*{8.3$\times$10$^{-11}$} & \multirow{2}*{2.1$\times$10$^{-5}$} & \multirow{2}*{8.0$\times$10$^{-9}$}\\ 
                 &            & [S\,II]\,$\lambda$6731 & ~95~($\pm$16) & 53 \\ 
                 &            &                                        &    & & & \\						        
\multirow{2}*{42.2 } & \multirow{2}*{2009 02 10} & [S\,II]\,$\lambda$6716 & 309~($\pm$33) & 67 & \multirow{2}*{5600} & \multirow{2}*{6.8$\times$10$^{-7}$} & \multirow{2}*{3.0$\times$10$^{-10}$} & \multirow{2}*{5.4$\times$10$^{-5}$} & \multirow{2}*{3.0$\times$10$^{-8}$}\\ 
                 &            & [S\,II]\,$\lambda$6731 & 139~($\pm$28) & 52\\ 
\hline
\end{tabular}
\tablefoot{
The listed entries are: slit position angle (PA), observing date,
detected FELs, spatial offsets of FELs in milli-arcsec measured by spectro-astrometry,
corresponding radial velocities $V_r$ of FELs, electron density $n_{\rm{e}}$ derived from the 
ratio [S\,II]\,$\lambda$6731/[S\,II]\,$\lambda$6716, line luminosity of [S\,II]\,$\lambda$6731, 
the mass outflow rate derived from [S\,II]\,$\lambda$6731 for an assumed outflow inclination angle of 
45$^{\circ}$, the line luminosity (of the red component) of O\,I\,$\lambda$8446, 
and the mass accretion rate based on $L_{8446}$.
}
\end{minipage}
\end{table*}

\begin{figure}[t]
\centering
\includegraphics[width=.8\linewidth,clip]{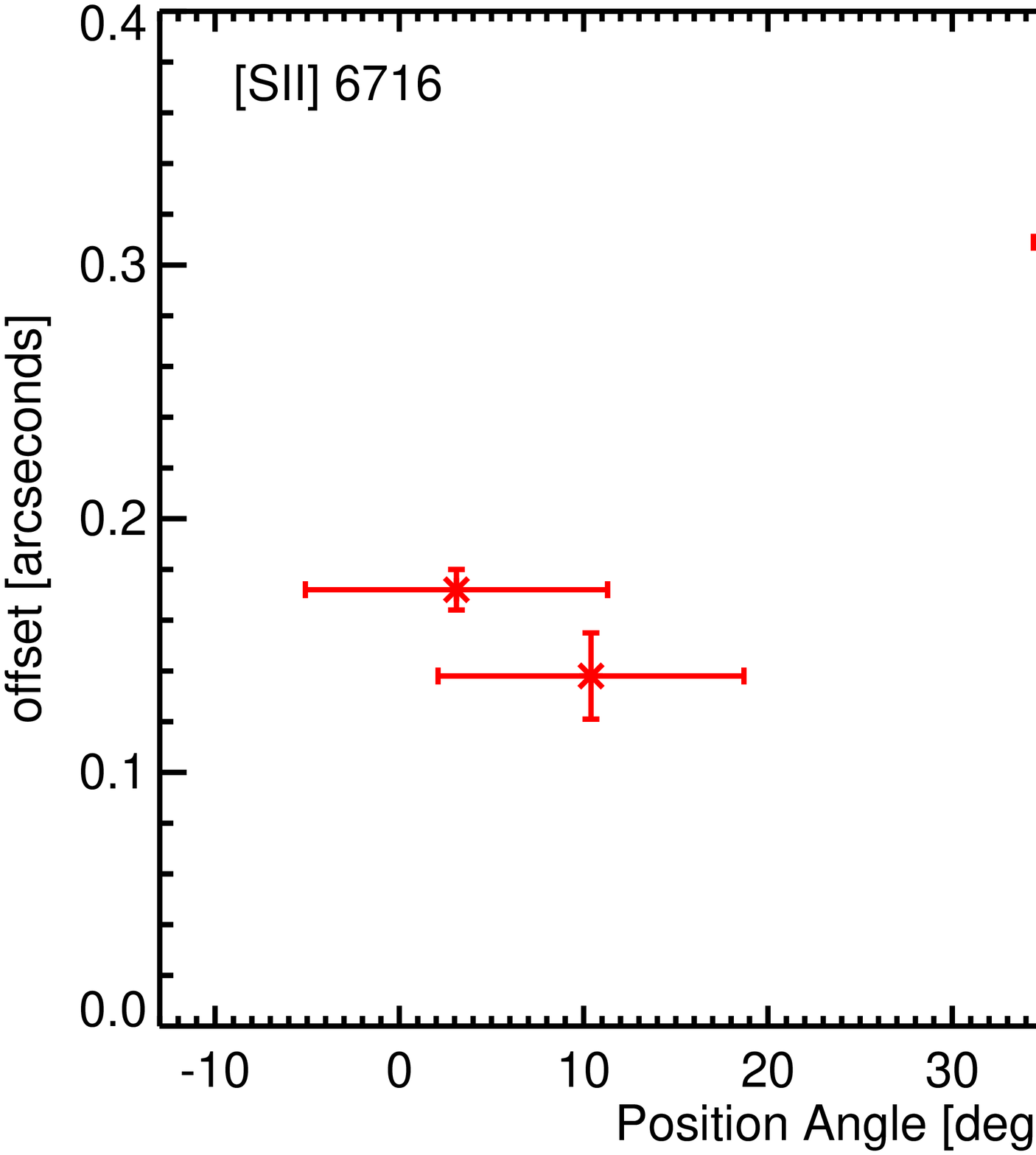}
\includegraphics[width=.8\linewidth,clip]{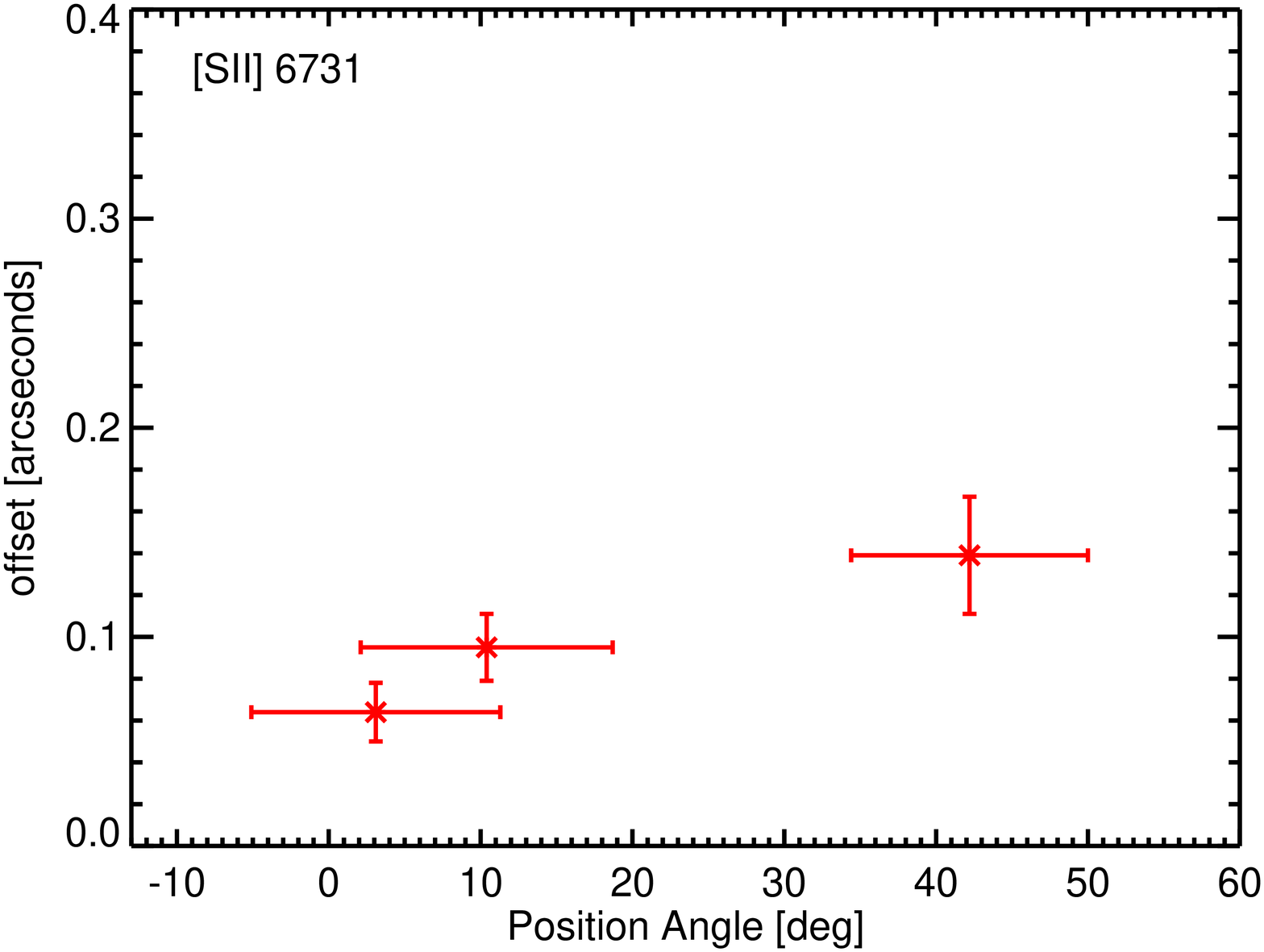}
\caption{
\label{fig:pa}
Spatial offsets of FEL of \iso143 as function of PA
for [S\,II]\,$\lambda$6716 (top panel) and [S\,II]\,$\lambda$6731 (bottom panel). 
}
\end{figure}

We have clearly discovered the spectro-astrometric signature of
an outflow in both [S\,II] lines at all three epochs in our UVES spectra of \iso143.
Figure\,\ref{fig:sa} shows in the middle and bottom rows
the line profiles of both the [S\,II]\,$\lambda\lambda$6716,6731 lines and 
their measured spatial offsets as a function of the velocity. The spatial offsets
are displayed for both the adjacent continuum and the continuum subtracted FEL.
The plotted errors in the spectro-astrometric plots are based on 1$\sigma$ errors in the 
Gaussian fit parameters.
Table\,\ref{tab:fel} lists the maximum spatial offsets and corresponding velocities 
of both [S\,II] lines in the order of increasing slit PA.
Offset errors in Table\,\ref{tab:fel} are based on the standard deviation in the continuum points.
For an overview and to constrain the outflow PA (see below), we plot the maximum spatial offsets 
as a function of the slit PA in Fig.\,\ref{fig:pa}.

Our spectro-astrometric analysis of the detected FELs of [S\,II] demonstrated
that they originate from spatially offset positions of
the continuum source by up to 200-300\,mas 
(about 30-50\,AU at the distance of Cha\,I)
at a velocity of up to 50-70\,km\,s$^{-1}$. 
We found the [S\,II]\,$\lambda$6716 emission to be 
spatially more extended than the [S\,II]\,$\lambda$6731 emission
(Table\,\ref{tab:fel}, Fig.\,\ref{fig:pa}),
which is consistent with the [S\,II]\,$\lambda$6716 line
tracing lower densities than the [S\,II]\,$\lambda$6731 line.

The spectra were taken at mean slit PAs of 3$^{\circ}\,\pm\,8^{\circ}$, 10$^{\circ}\,\pm\,8^{\circ}$, 
and 42$^{\circ}\,\pm\,8^{\circ}$ (cf. Sect.\,\ref{sect:obs}).
As illustrated in Fig.\,\ref{fig:pa}, we see
a tendency for the estimated outflow extension to increase with increasing slit PA.
The data thus hint that by varying the slit PA from $3^{\circ}$ to $42^{\circ}$, 
one approaches the actual outflow PA.
Follow-up observations, ideally spectra taken at two orthogonal slit PAs,
are required to further constrain the outflow PA of \iso143, in particular since the S/N is 
moderate for the spectra at 42$^{\circ}$.

\subsection{Electron density and mass outflow rate}
\label{sect:mout}

The line intensity ratio [S\,II]$\lambda$6731/[S\,II]$\lambda$6716
can be used to derive the gas electron density $n_{\rm{e}}$ in the line-emitting region
because these two lines have nearly the same excitation
energy and, therefore, their relative excitation rates only depend on the ratio of collisional
strengths, i.e. on the density
(Osterbrock \& Ferland 2006, p. 121 ff; Bacciotti et al. 1995).
We measured an [S\,II] line ratio for \iso143 of 2.0 
in spectra from Jan. 2009 (slit PA=3$^{\circ}$) and Feb. 2009 (42$^{\circ}$, cf. Table\,\ref{tab:ew}) 
indicating
an electron density $n_{\rm{e}}$ of about 5600\,cm$^{-3}$.
For the epoch Mar. 2008 (10$^{\circ}$), the estimated [S\,II] line ratio indicates an $n_{\rm{e}}$ value
above the line critical density $n_c$, 
thus only a lower limit
$n_{\rm{e}} > n_{\rm{c}} \sim 2 \times 10^{4}$\,cm$^{-3}$ can be derived in this case.

We proceeded to estimate the mass loss rate of \iso143
following Hartigan et al. (1995; cf. also Comer\'on et al. 2003).
For this purpose we derived the 
line luminosity of [S\,II]$\lambda$6731 from its measured EW (Table\,\ref{tab:ew})
and the dereddened $R_0$ magnitude:
\begin{equation}
\label{eq:lum}
L(\lsun) = 6.71 \times 10^{-5} \, d^2 \, (pc) \, EW(\AA) \, 10^{-0.4 R_0} \quad .
\end{equation}
We used a distance $d$ of 162.5\,pc (Luhman 2007), an R-band magnitude of 18.02\,mag, which is obtained in the 
WFI R-band filter (Lopez-Marti et al. 2004), and 
an extinction of A$_{R(WFI)}$=2.97\,mag. The latter was converted from $A_J$=1.13 (Luhman 2007) by interpolating 
the reddening law of Mathis (1990) for $R_V$=5.0 (Luhman 2004).
The line luminosity we estimated for [S\,II]$\lambda$6731 in \iso143 
is 7--9$\times$10$^{-7}$ $\lsun$ (Table\,\ref{tab:fel}).

The mass loss rate was estimated from the line luminosity $L_{6731}$ for  $n_e < n_c$ 
by the following equation (Comer\'on et al. 2003): 
\begin{equation}
\label{eq:mout1}
\dot{M}_{out} (\mperyr) = 4.51 \times 10^9 \, \left( 1+ \frac{n_c}{n_e} \right) \times \frac{L_{6731}(\lsun) \, V_{\bot} (km\,s^{-1})}{l_{\bot} (cm)}
\end{equation}
where $n_c$ is the critical density and $V_{\bot}$ and $l_{\bot}$ are the estimated tangential flow 
speed and scale length of the emission.
In the case of the [SII]$\lambda$6731 line that is within the high density
limit (Mar. 2008, 10$^{\circ}$), the mass loss rate does not depend upon the electron density
(Hartigan et al. 1995) and Eq.\,\ref{eq:mout1} becomes
\begin{equation}
\label{eq:mout2}
\dot{M}_{out} (\mperyr) = 4.51 \times 10^9 \times \frac{L_{6731}(\lsun) \, V_{\bot} (km\,s^{-1})}{l_{\bot} (cm)} \quad .
\end{equation}
For the scale length of the emitting region $l_{\bot}$, 
we adopted the size of the aperture (cf. e.g. Hartigan et al. 1995), 
i.e. the UVES slit width of the observations of 1$^{\prime\prime}$. 
For the outflow tangential velocity $V_{\bot}$, we exploited the kinematic properties determined 
from spectro-astrometry of the [SII]$\lambda$6731 line (Table\,\ref{tab:fel}),
which indicate a radial velocity $V_r$ of the outflow of \iso143 of 
around 50\,km\,s$^{-1}$. 
Although the outflow inclination is unknown in principle, we can roughly estimate the tangential velocity $V_{\bot}$
from the observed $V_r$ by assuming that the outflow direction is perpendicular to the plane of the 
accretion disk and using the disk inclination value derived from the SED fitting 
(Harvey et al. 2012b; Y. Liu, pers. comm.).
The disk inclination is, unfortunately, not very well constrained, 
because of the degeneracy of the SED model; 
the best-fit value is close to face-on (15$^{\circ}$-25$^{\circ}$), but disk inclination values 
0$^{\circ}$--50$^{\circ}$ are almost equally likely
(cf. Sect.\ref{sect:iso143}).
The visibility of the red outflow component indicates an at least not completely face-on viewing angle of the disk, and 
we therefore adopt, somewhat arbitrarily, an outflow inclination angle of 45$^{\circ}$ 
for calculating the mass outflow rate, i.e. $V_r$=$V_{\bot}$. 

The estimated outflow rate for \iso143 
(0.8--3$\times 10^{-10}$\,$\mperyr$, Table\,\ref{tab:fel}) is consistent with 
previously determined mass loss rates of brown dwarfs and VLMS (Whelan et al. 2009a; Bacciotti et al. 2011).
We note that assuming a larger outflow inclination angle, say of 65$^{\circ}$ (corresponding to a 
disk inclination of 25$^{\circ}$), would result in outflow rates that are
a factor of two lower than the given values.
We find the same outflow rate for the observations at a slit PA of 3$^{\circ}$ and 42$^{\circ}$
indicating no significant change as a function of slit PA from these data.
While we derive a smaller outflow rate for observations at 10$^{\circ}$,
this deviation in the outflow rate by a factor of about four
can be traced back to a small difference in the EW value of 0.1\,{\AA} and is, therefore,
not taken as sound proof of a variable mass loss rate.
Interestingly, a lower mass loss rate in this epoch would correlate with a decreased mass accretion rate in this epoch 
(see Sect.\,\ref{sect:macc}).

\subsection{Mass accretion rate}
\label{sect:macc}

The mass accretion rate of \iso143 was derived from the accretion luminosity, 
\begin{equation}
\label{eq:macc}
\dot{M}_{acc} = \left( 1 - \frac{R_*}{R_{in}} \right ) ^{-1} \frac{L_{acc} R_*}{G M_*} \sim 1.25 ~\frac{L_{acc} R_*}{G M_*}
\end{equation}
(Gullbring et al. 1998), by the use of the OI\,$\lambda8446$ emission line
as an indirect accretion tracer. We can assume that this line 
is predominantly produced by accretion, possibly except for the blue component we find in the profile 
of Mar. 2008 (cf. Sect.\,\ref{sect:lines}).
We determined the line luminosity of OI\,$\lambda8446$ (for the spectrum of Mar. 2008 only the line 
luminosity of the 
red component) by applying Eq.\,\ref{eq:lum} and using the dereddened $i$-band magnitude.
We adopted an $i$-band magnitude of 15.51\,mag (Luhman 2004), which was obtained in the 
DENIS $i$-band filter, and an extinction of A$_{i(DENIS)}$=2.37\,mag, which is
again converted from $A_J$=1.13 (Luhman 2007) by interpolating
the extinction law of Mathis (1990).
A comparison of the determined $L_{8446}$ (see Table\,\ref{tab:fel}) with an empirical correlation of
line and accretion luminosity for OI\,$\lambda8446$ (Herczeg \& Hillenbrand 2008) allowed us to
roughly estimate the accretion luminosity $L_{acc}$.
This accretion luminosity is converted into a 
mass accretion rate by means of Eq.\,\ref{eq:macc} by adopting
a model-dependent mass of 0.18\,$\msun$ and
a radius determined from the Stefan-Boltzmann law of 1.01\,$\rsun$ 
(cf. Sect.\,\ref{sect:iso143}).

We estimate a mass accretion rate for \iso143 between 8$\times 10^{-9}$
(Mar. 2008) and 2-3$\times 10^{-8}$\,$\mperyr$
(Jan. and Feb. 2009, Table\,\ref{tab:fel}).
These mass accretion rates are close to what was found for other VLMS, although at the higher edge of 
this distribution
(e.g., Fang et al. 2009; Herczeg et al. 2009),
but given the large scatter in the empirical $L_{line}$-$L_{acc}$ correlation they are 
consistent.
We can also convert the EW of H$\alpha$ measured for \iso143 (118\,{\AA}, Luhman 2004)
into a mass accretion rate by using the empirical relation of Fang et al. (2009) and assuming 
that the H$\alpha$ emission is entirely related to accretion. Following this path, we find a 
mass accretion rate of 1$\times 10^{-9}$\,$\mperyr$, which is about an order of magnitude less
than the value derived using the O\,I line. 

With the caveat in mind that accretion rate determinations are affected by large uncertainties, 
we find indications of a variable accretion rate of \iso143 based on the O\,I line of 0.6\,dex. This 
is consistent with other studies monitoring accretion rates of CTTS and that include VLMS, 
e.g. that of Costigan et al. (2012) who find variability of 0.4-0.9\,dex for a sample of G2-M5.75 stars 
(cf. also Scholz \& Jayawardhana 2006).
The observation of accretion in \iso143 derived from the O\,I line being on a lower level in Mar. 2008
compared to Jan. and Feb. 2009 is also in line with
the observations of the red component of the Ca\,II$\lambda$8498 line, which we attribute  
to accretion (cf. Sect.\,\ref{sect:lines} and Fig.\,\ref{fig:lp}).

Tentative hints were found of a relatively high ratio of mass-outflow to mass-accretion rate 
for brown dwarfs and VLMS
($\geq$40\,\%, Comer\'on et al. 2003; Whelan et al. 2009; Bacciotti et al. 2011)
compared to that of CTTS (about 1--10\,\%, e.g. Ray et al. 2007). 
The $\dot{M}_{out}/\dot{M}_{acc}$ ratio we derive for \iso143 appears to be 
closer to the CTTS value, as it is about 1\,\%
using the O\,I-based accretion rate 
and about 20\,\% using the H$\alpha$-based accretion rate.


\section{Conclusions}
\label{sect:concl}

We have discovered that the young very low-mass star \iso143 (M5)
is driving an outflow based on spectro-astrometry 
of forbidden [S\,II] lines in UVES/VLT spectra.
This adds another outflow discovery to the very low-mass regime (M5-M8),
where only about a handful of outflows have been confirmed so far.
We have demonstrated that the forbidden [S\,II] emission in \iso143 is formed spatially offset 
from the central source by up to 200-300\,mas 
(about 30-50\,AU at the distance of Cha\,I) at a velocity of up to 50-70\,km\,s$^{-1}$. 
We found this outflow to be intrinsically asymmetric as 
only the red outflow component is visible in the [S\,II] lines.
This indicates an asymmetry in the launched outflow itself and/or
in the distribution of gas in the immediate surrounding.
Furthermore, in both cases there appears to be no or only little obscuration by the 
circumstellar accretion disk despite a low or intermediate disk inclination angle ($<60^{\circ}$) 
and a relatively flat disk geometry (Harvey et al. 2012b; Y. Liu, pers. comm.).
To our knowledge \iso143 is only the third detected T~Tauri object showing a
stronger red outflow component in spectro-astrometry, 
following RW\,Aur (G5, Hirth et al. 1994b) and \iso217 (M6.25, 
Whelan et al. 2009a; Joergens et al. 2012), although in general asymmetric jets are
common for CTTS ($\sim$50\,\%, Hirth et al. 1994b, 1997).
We have shown here that, including \iso143,
two out of seven outflows confirmed in the very low-mass regime (M5-M8) are intrinsically 
asymmetric.
We estimated the mass outflow rate of \iso143 based on the [S\,II] lines
to 0.8--3$\times 10^{-10}$\,$\mperyr$, which is consistent with 
previously determined mass-loss rates of brown dwarfs and VLMS
(Whelan et al. 2009a; Bacciotti et al. 2011).

We furthermore detected line emission of \iso143 from several activity-related 
emission lines of Ca\,II, O\,I, He\,I, and [Fe\,II].
Based on a line profile analysis we showed that likely formation sites of these emission lines include
the chromosphere, the accretion disk, the infall zone, and a wind.
A decomposition of the Ca\,II\,$\lambda$8498 IR line revealed three emission components,
which can be attributed to narrow-band chromospheric emission,
broader blue-shifted emission of a variable wind, and very broad red-shifted component
probably produced in the magnetospheric infall zone of the system.
The detected broad O\,I\,$\lambda$8446 emission appears
to be mainly formed in accretion-related processes in the disk. In addition, we
observed a blue-shifted signature in the O\,I line at one epoch that is 
tracing the wind signature seen in the Ca\,II line suggesting a common formation site in a variable wind.
The origin of the He\,I\,$\lambda$7065 emission can be attributed to the chromosphere
or the transition zone between the chromosphere and the corona
based on its narrow profile that is centered on the stars' velocity and on the high excitation
temperature of this line transition.
We estimated the mass accretion rate based on the O\,I$\lambda$8446 line to 
0.8-3$\times 10^{-8}$ and based on the H$\alpha$ EW from the literature (Luhman 2004)
to 1$\times 10^{-9}$\,$\mperyr$, which is consistent with that of other VLMS
(e.g., Fang et al. 2009; Herczeg et al. 2009).
The derived ratio of the mass-outflow to mass-accretion rate 
$\dot{M}_{out}/\dot{M}_{acc}$ of \iso143 of 1-20\,\% indicates
a value close to that of other CTTS (about 1--10\,\%, e.g. Ray et al. 2007)
and appears to not support previous 
tentative findings of a very high $\dot{M}_{out}/\dot{M}_{acc}$ ratio
for brown dwarfs and VLMS
($\geq$40\,\%, Comer\'on et al. 2003; Whelan et al. 2009; Bacciotti et al. 2011).

We conclude that the very low-mass star \iso143 exhibits 
the well known activity features seen for higher mass T~Tauri stars, 
such as chromospheric activity, accretion, magnetospheric infall, winds, and an outflow
providing further confirmation that the T~Tauri phase continues at very low masses.

\begin{acknowledgements}
We thank the ESO staff at Paranal for the execution of the observations
presented here in service mode. 
We are grateful to Y. Liu for providing details on the disk properties of \iso143
and to I. Pascucci, Y. Liu, and S. Wolf for
fruitful discussions on the topic of this paper. 
Furthermore, we would like to thank an anonymous referee for helpful comments
that allowed us to significantly improve the paper.
Part of this work was funded by the ESF in Baden-W\"urttemberg.
\end{acknowledgements}

\end{document}